\documentclass[twocolumn,pra,aps,superscriptaddress,showpacs]{revtex4-2}
\usepackage{amsmath}
\usepackage{esint}
\usepackage{braket}
\usepackage{graphicx}
\usepackage{bbm}
\usepackage{subfigure}
\usepackage{pifont}
\usepackage{csquotes}
\usepackage{float}
\usepackage{amsmath}
\usepackage{amsfonts}
\usepackage{graphicx}
\usepackage{bm}
\usepackage{dcolumn}
\usepackage{natbib}
\usepackage{color}
\usepackage{textcomp}
\usepackage[colorlinks,citecolor=blue,linkcolor=red]{hyperref}
\usepackage{graphicx}% Include figure files
\usepackage{dcolumn}% Align table columns on decimal point
\usepackage{bm}% bold math
\usepackage{bbm}
\usepackage[normalem]{ulem}
\usepackage{xcolor}
\makeatletter
\@ifundefined{textcolor}{}
{%
	\definecolor{BLACK}{gray}{0}
	\definecolor{WHITE}{gray}{1}
	\definecolor{RED}{rgb}{1,0,0}
	\definecolor{GREEN}{rgb}{0,1,0}
	\definecolor{BLUE}{rgb}{0,0,1}
	\definecolor{CYAN}{cmyk}{1,0,0,0}
	\definecolor{MAGENTA}{cmyk}{0,1,0,0}
	\definecolor{YELLOW}{cmyk}{0,0,1,0}
}

\usepackage{bm}

\makeatother

\begin{document}
	
	% 标题与作者
	\title{Boundary Time Crystals Beyond Mean-Field Theory: A Stroboscopic Rotating-Wave Approximation}
	\author{Zeping Liu}
	\affiliation{Zhejiang Key Laboratory of Quantum State Control and Optical Field Manipulation, Department of Physics, Zhejiang Sci-Tech University, 310018 Hangzhou, China}
	\author{Yaotian Li}
	\affiliation{Zhejiang Key Laboratory of Quantum State Control and Optical Field Manipulation, Department of Physics, Zhejiang Sci-Tech University, 310018 Hangzhou, China}
	\author{Zhaoyu Fei}
	\email{feizhaoyu@zstu.edu.cn}
	\affiliation{Zhejiang Key Laboratory of Quantum State Control and Optical Field Manipulation, Department of Physics, Zhejiang Sci-Tech University, 310018 Hangzhou, China}
	\author{Xiaoguang Wang}
%	\email{xgwang@zstu.edu.cn}
	\affiliation{Zhejiang Key Laboratory of Quantum State Control and Optical Field Manipulation, Department of Physics, Zhejiang Sci-Tech University, 310018 Hangzhou, China}
	
	\date{\today}
	
	\begin{abstract}
		Boundary time crystals are a class of exotic dissipative quantum phases that spontaneously break continuous time-translation symmetry in the thermodynamic limit of open quantum systems. In finite-size systems, the long-time evolution of boundary time crystals exhibit decaying oscillations that cannot be captured by widely used mean-field theory. To address this issue, we develop an effective approach called the stroboscopic rotating wave approximation, which provides a well approximate state for the long-time evolution of boundary time crystals in the strongly driven regime. 
		In this approach, the order parameter exhibits both a long-time decaying envelope governed by an effective Lindblad superoperator and short-time oscillations dominated by a reduced quantum dynamical semigroup. 
		Our results reveal that the competition among dephasing processes along three directions induces persistent oscillations, marking the emergence of the boundary-time-crystal phase. 
		We obtain the analytical expressions for the steady-state density operator, the oscillation period and the decay rate of the order parameter when the coherent energy splitting exceeds the dissipation rate. 
		Our work provides a beyond-mean-field tool for studying the dynamics of periodically driven open quantum systems and understanding the formation of time crystals.
	\end{abstract}
	
	\maketitle
	
	% ====== 正文 ======
	\section{Introduction}
	
	When quantum systems undergo spontaneous breaking of time-translation symmetry, they give rise to the so-called time crystals~\cite{sacha2017time,khemani2019brief,sacha2020time}. The concept was first proposed by Wilczek in 2012~\cite{PhysRevLett.109.160401}, in analogy with the spontaneous breaking of spatial translation symmetry in conventional crystals. 
	The time crystals exhibit periodic oscillations even in quantum ground state.	 
	However, a no-go theorem subsequently ruled out the existence of time crystals in the ground states of general Hamiltonians (or at the thermal equilibrium)~\cite{watanabe2015absence,bruno2013impossibility}. 
	Subsequent studies have revealed two types of time crystals: discrete time crystals and dissipative time crystals. Discrete time crystals break discrete time-translation symmetry, characterized by persistent macroscopic oscillations at integer~\cite{else2020discrete,else2016floquet,lazarides2020time,kosior2018dynamical,yao2017discrete,riera2020time} (or fractional~\cite{PhysRevA.99.033626,article}) multiples of the driving period, which have recently been confirmed experimentally~\cite{gong2018discrete,zhang2017observation,kongkhambut2022observation,kyprianidis2021observation,smits2018observation,rovny2018observation,liu2024higher,zhu2019dicke,randall2021many,bar2024discrete}. Dissipative time crystals, on the other hand, break continuous time-translation symmetry in open quantum systems~\cite{PhysRevLett.121.035301,buvca2019non,kessler2019emergent,huang2025observation,krishna2023measurement,passarelli2022dissipative,makinen2025continuous,russo2025quantum,li2024time,wang2025boundary,yang2025emergent}. Among these studies, a particularly notable subclass is the boundary time crystal (BTC) , which emerges at the system’s boundary and whose oscillation period is determined by the coupling constant and the driving frequency~\cite{montenegro2023quantum}. 
	These features can be exploited to achieve higher measurement precision and greater robustness in quantum metrology and quantum sensing~\cite{montenegro2023quantum,cabot2024continuous}.
	
	In the thermodynamic limit, the dynamics of BTCs can be well described by mean-field theory (MFT), which successfully identified the corresponding nonlinear dynamical equations, the phase-transition point, and the critical exponents of BTCs~\cite{barberena2024critical,PhysRevB.104.014307,PhysRevA.105.L040202,montenegro2023quantum,PhysRevA.110.012220}.
	However, in numerical simulations and experiments, the thermodynamic limit is unattainable, and finite-size effects are inevitable. 
	For finite-size systems, by employing the error function proposed in Ref.~\cite{PhysRevLett.121.035301}, we demonstrate that in the BTC phase, the errors in the system's dynamical evolution introduced by the MFT grow exponentially, which renders it entirely invalid.
	A unified and practically useful analytical framework that can reliably capture the long-time dynamics of BTC in finite-size systems is still absent. 
	To approximate the long-time evolution of BTC in finite-size systems, we propose a method called the stroboscopic rotating wave approximation (SRWA).
				
	Previous studies on SRWA have mainly focused on the Schrödinger equations, where the approximate dynamics typically manifest at the stroboscopic points, i.e., the integer multiples of driving period~\cite{zeuch2020exact,blanes2009magnus,minganti2022arnoldi,restrepo2016driven}. In our work, we extend the framework to open quantum systems. The method divides the full-time dynamics into two parts: (i) the long-time decay at the stroboscopic points, governed by an effective Lindblad superoperator; (ii) the short-time oscillations between adjacent stroboscopic points, described by a reduced quantum dynamical semigroup.
	The results reveal that the competition among dephasing processes along three directions induces persistent oscillations, marking the emergence of the boundary-time-crystal phase.
	Within this framework, we derive analytical expressions for the steady-state density matrix, the oscillation period and the decay rate of the order parameter which go beyond those obtained by using the MFT.
	Our work establishes an effective dynamical approximation tool for periodically driven open quantum systems and deepens the understanding of the formation mechanism of time crystals.
	
	The remainder of this paper is organized as follows.
	In Sec.~\ref{sec:model}, we introduce the general framework of boundary time crystals and analyze the breakdown of the MFT in finite-sized systems.
	In Sec.~\ref{sec:method}, we present the SRWA, providing explicit expressions for the effective Lindblad superoperator and the reduced quantum dynamical semigroup.
	In Sec.~\ref{sec:results}, we construct the full-time dynamics of BTC using SRWA and obtain the analytical expressions for the steady-state density operator, the oscillation period and the decay rate of the order parameter in the strongly driving regime where the coherent energy splitting exceeds the dissipation rate.
	In Sec.~\ref{sec:new}, we adopt an alternative rotation frequency to improve the accuracy of the SRWA for large system sizes, and provide an error analysis of it.
	Finally, in Sec.~\ref{sec:discussion and conclusion}, we summarize the main findings, discuss their physical implications, and outline potential directions for future researches.
	
	\section{Boundary Time Crystal in Finite-Sized Systems}
	\label{sec:model}	% 描述体系 Hamiltonian、Lindblad 主方程、符号定义等
	We consider a system composed of \( N \)  spin-1/2 particles, and the total spin is \( S = N/2 \). The collective angular momentum operators are 
	$
	\hat{S}_\alpha = \frac{1}{2} \sum_{j} \hat{\sigma}_\alpha^{(j)},  \alpha = x, y, z,
	$
	where \( \hat{\sigma}_\alpha^{(j)} \) denotes the Pauli matrix of the $j$ th spin. 
	The Hamiltonian is taken as
	\begin{equation*}
		\hat{H} = \omega_{0} \hat{S}_x,
	\end{equation*}
	where \( \omega_{0} \) denotes the single-particle coherent energy splitting~\cite{PhysRevLett.121.035301}, and $T_{0}=2\pi /\omega_{0}$ represents the driving period. These particles represent the boundary of a \(d\)-dimensional quantum many-body system.	Its time evolution is governed by an effective master equation, including collective spin dissipation due to the interaction with the bulk system:
	\begin{equation}
		\frac{\mathrm{d} \tilde{\rho}}{\mathrm{d} t} = -i[\hat{H}, \hat{\rho}] +  \frac{\kappa}{N}  \left(  2 \hat{S}_- \hat{\rho} \hat{S}_+ -  \hat{S}_+ \hat{S}_- \hat{\rho} -  \hat{\rho} \hat{S}_+ \hat{S}_-   \right),
		\label{eq:1}
	\end{equation}
    where $\kappa$ is the dissipation rate, and the operators $\hat{S}_{\pm} = \hat{S}_{x} \pm i \hat{S}_{y}$ denote the raising and lowering operators, respectively. 
    Here, $\hat{\rho}(t)$ represents the reduced density matrix of the boundary subsystem, obtained by tracing out the bulk degrees of freedom from the full many-body state. For simplicity, the explicit time dependence $t$ will be omitted in what follows.
	The interplay between coherent driving and collective dissipation gives rise to two distinct dynamical phases. For the static phase, $\omega_{0}/\kappa < 1$,  the dissipation term on the right-hand side primarily governs the evolution, causing the system to rapidly relax to a steady state. In the opposite case, when $\omega_{0}/\kappa > 1$, the system exhibits a persistent oscillation under strong driving, indicating the emergence of the BTC phase. Meanwhile, the Liouvillian spectrum beacomes gapless with a nonzero imaginary part. 	 
	
\begin{figure*}[t]
		\renewcommand{\figurename}{FIG.}
		\centering
		
		% --- Row 1 ---
		\begin{minipage}{0.49\linewidth}
			\centering
			\includegraphics[width=\linewidth]{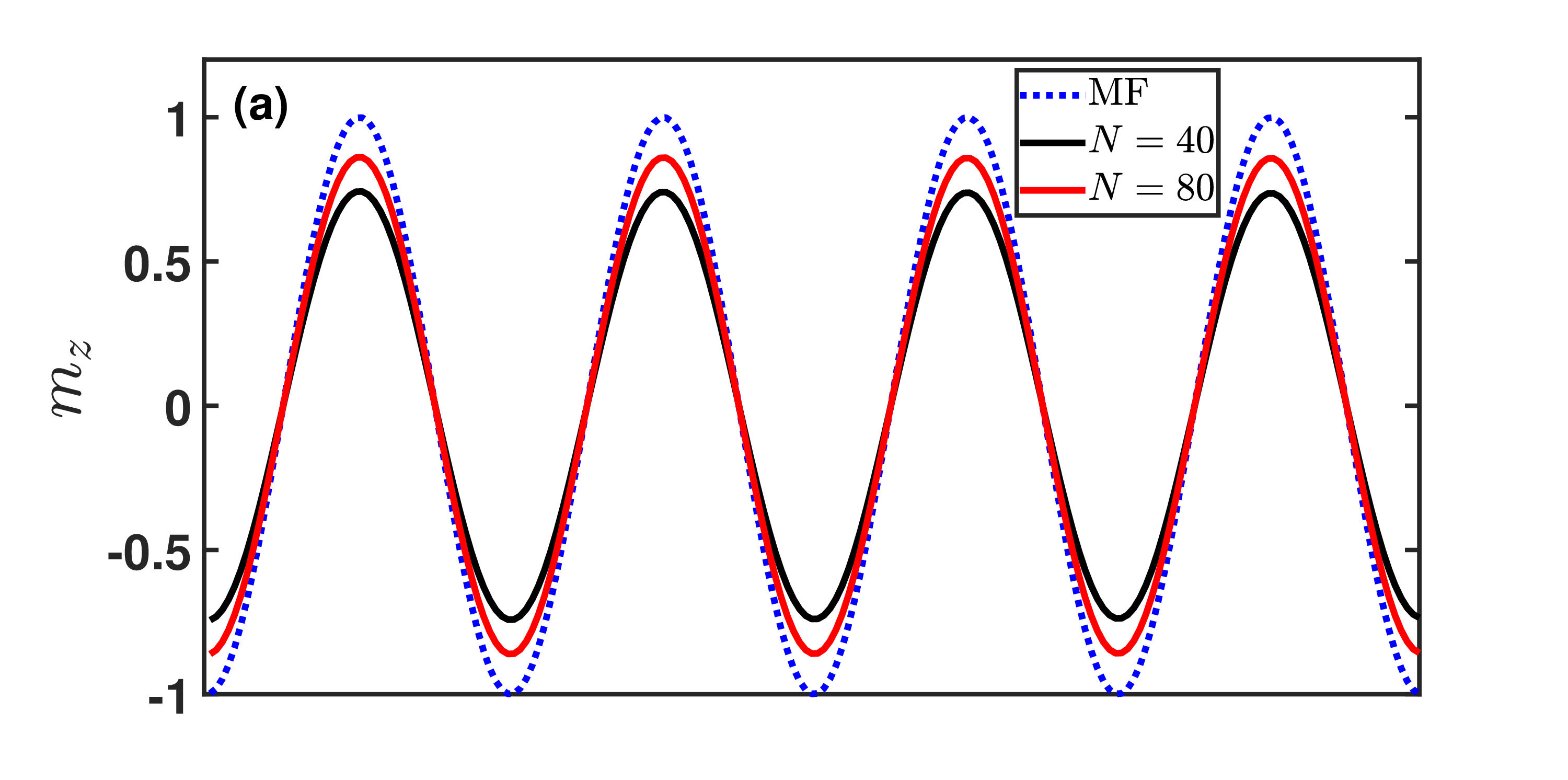}
		\end{minipage}
		\hspace{2pt}
		\begin{minipage}{0.49\linewidth}
			\centering
			\includegraphics[width=\linewidth]{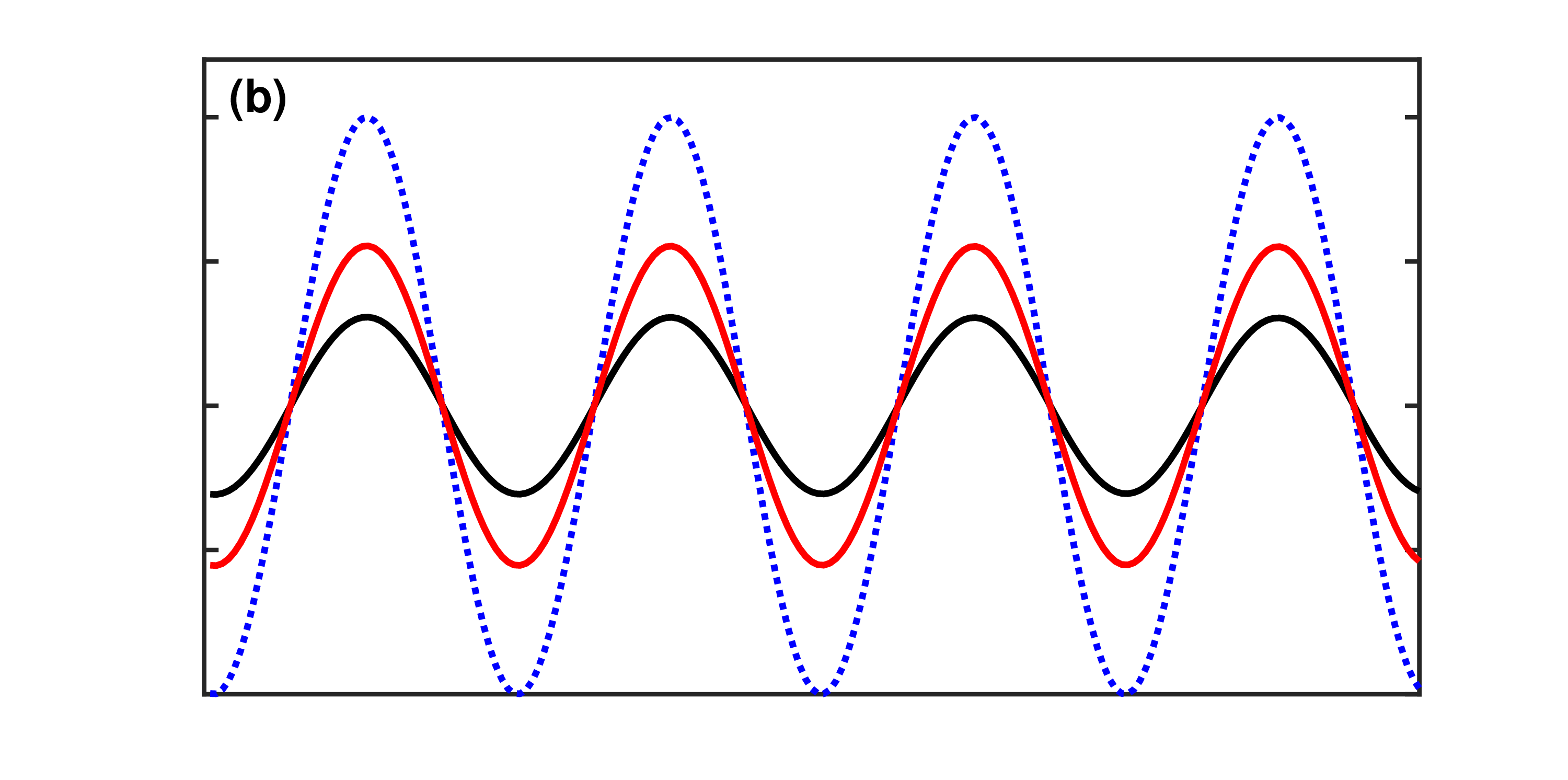}
		\end{minipage}
		
		\vspace{4pt}
		
		% --- Row 2 ---
		\begin{minipage}{0.49\linewidth}
			\centering
			\includegraphics[width=\linewidth]{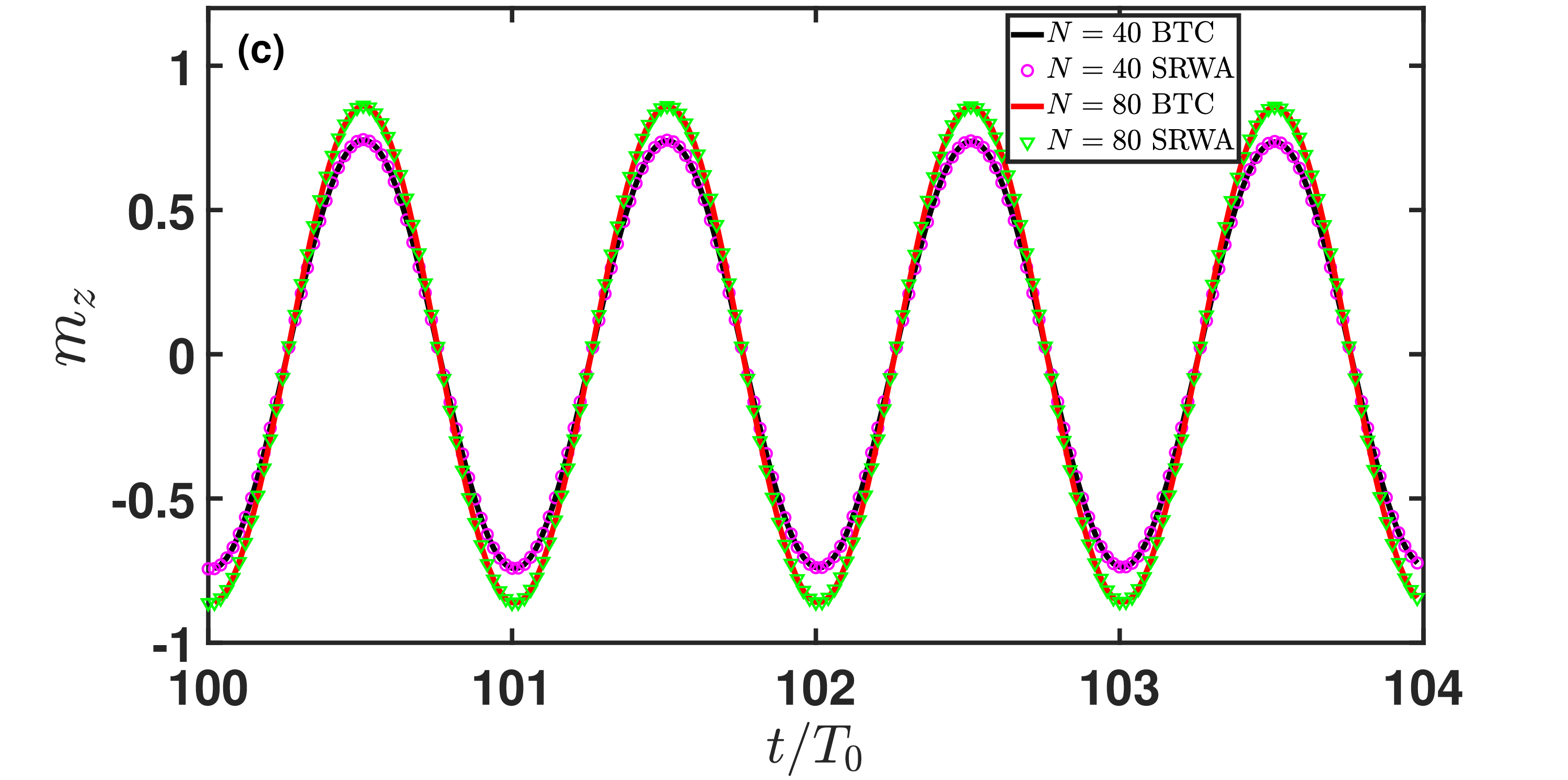}
		\end{minipage}
		\hspace{2pt}
		\begin{minipage}{0.49\linewidth}
			\centering
			\includegraphics[width=\linewidth]{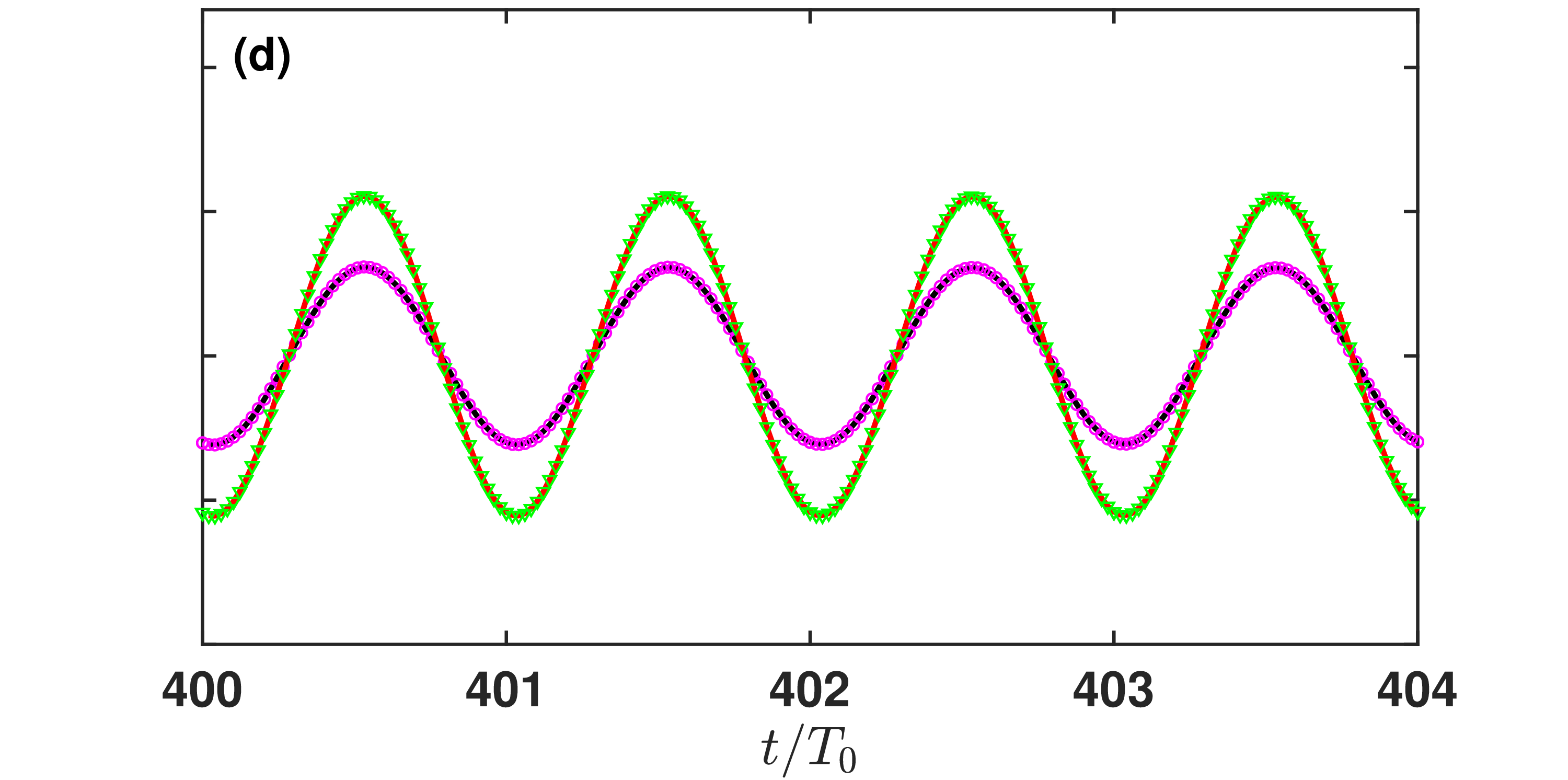}
		\end{minipage}
		
		% --- Row 3 ---
		\begin{minipage}{0.49\linewidth}
			\centering
			\includegraphics[width=\linewidth]{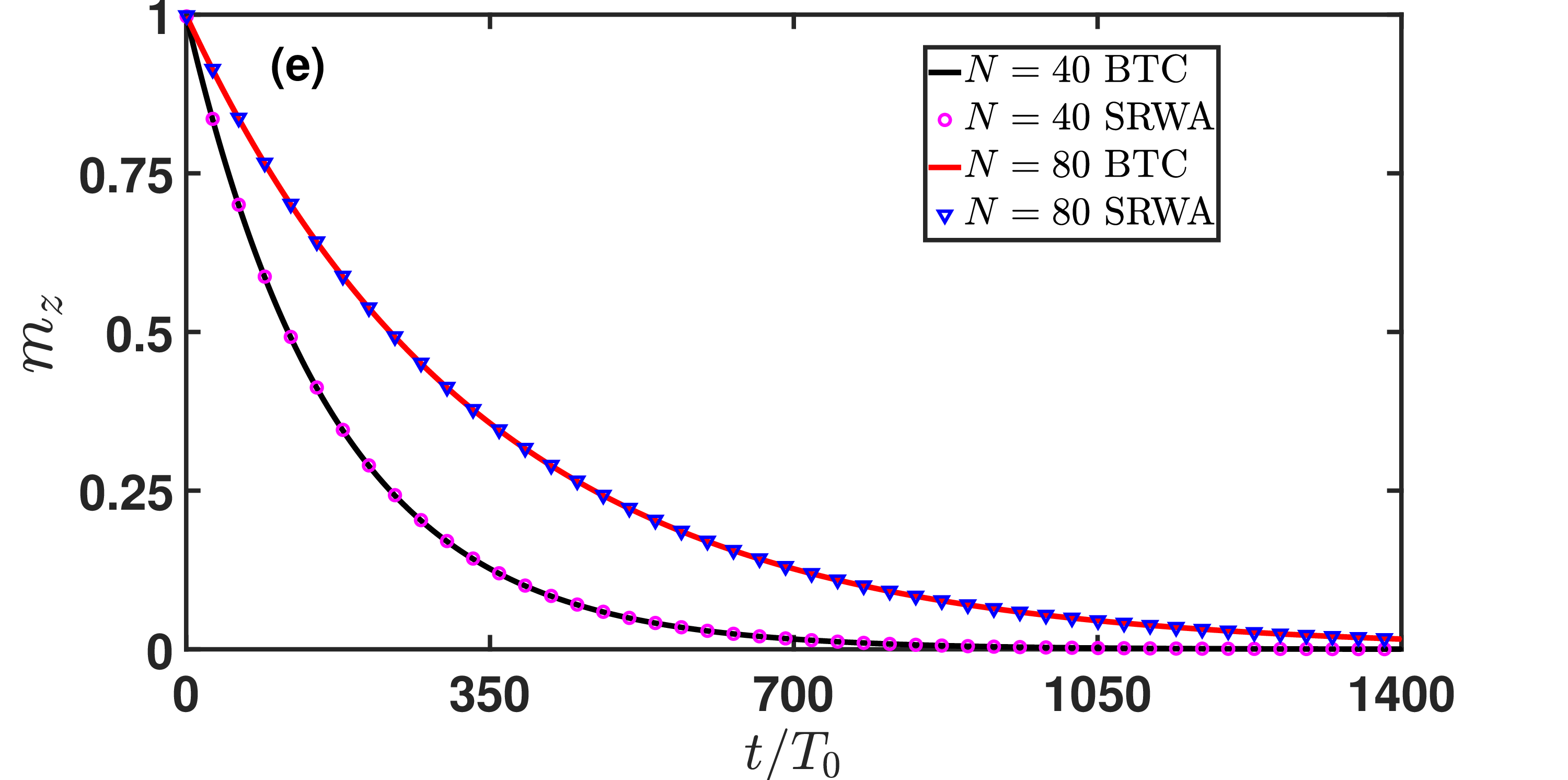}
		\end{minipage}
		\hspace{2pt}
		\begin{minipage}{0.49\linewidth}
			\centering
			\includegraphics[width=\linewidth]{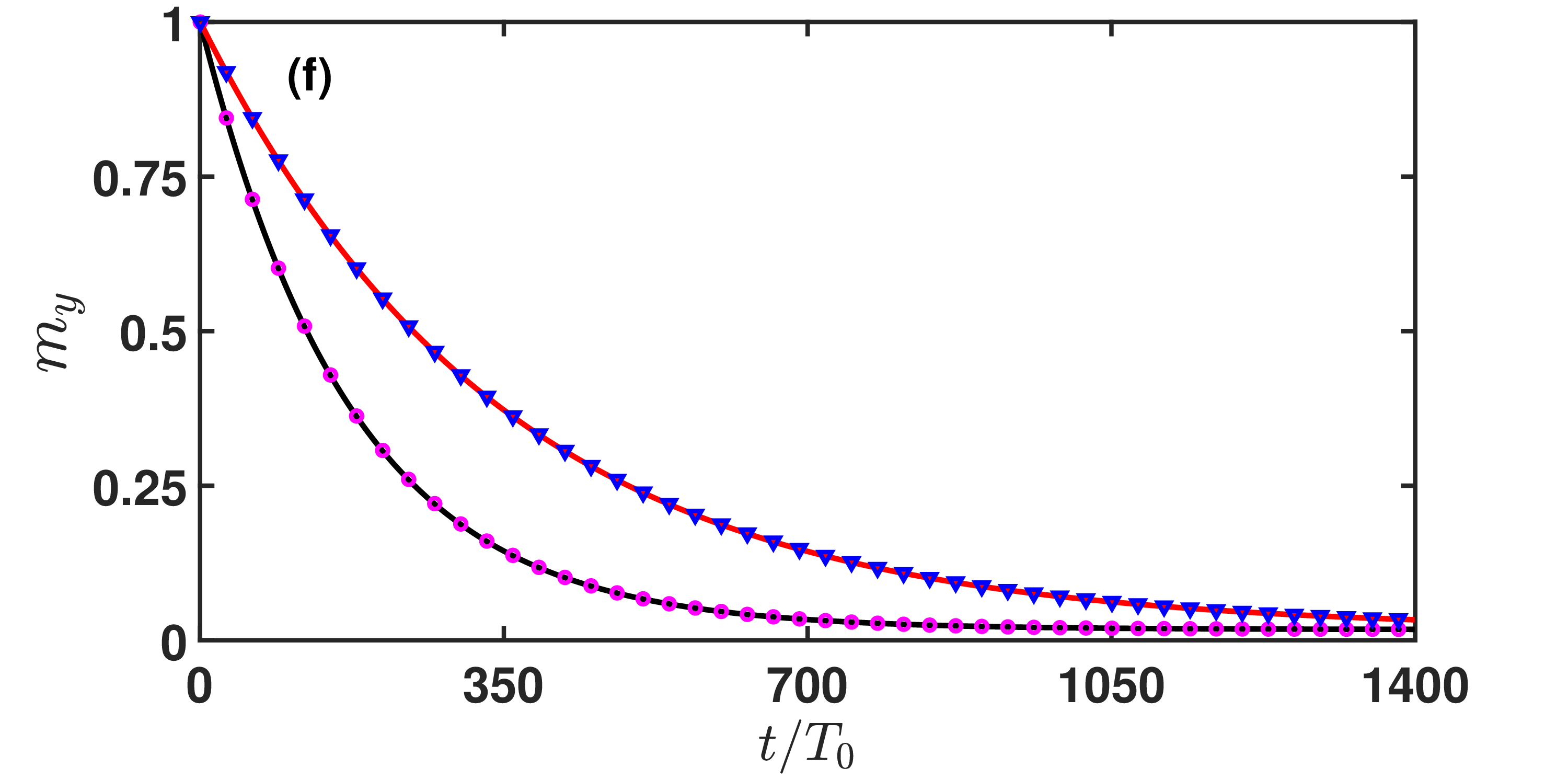}
		\end{minipage}
		
		\caption{
			Time evolution of the order parameter $m_z$ in the BTC.
			The initial state for all panels is chosen as all spins pointing down, and $\omega_0 / \kappa = 80$.
			The time evolutions are evaluated over the driving-period intervals $100$--$104$ and $400$--$404$,
			corresponding to panels (a, c) and (b, d), respectively.
			The blue dashed lines represent the results obtained from the second-order cumulant mean-field approximation~\cite{PhysRevA.110.012220},
			while the black and red solid lines represent the results obtained from the numerical simulation of Eq.~\eqref{eq:1}
			for system sizes $N=40$ and $N=80$, respectively.
			The purple circles and green inverted triangle represent the results obtained by using the SRWA.
			In panels (e, f), the black and red solid lines show the envelopes of $m_z$ and $m_y$ over 1400 driving periods,
			extracted from the numerical simulation of Eq.~\eqref{eq:1} for $N=40$ and $N=80$, respectively,
			while the purple circles and blue inverted triangle show the results obtained by using the SRWA .
		}
		\label{fig1}
	\end{figure*}

	Theoretically, the BTC phase can be well described using the MFT~\cite{PhysRevA.105.L040202, PhysRevB.104.014307, barberena2024critical, carollo2024applicability}. In the thermodynamic limit, we define the magnetization 
	$m_{\alpha} \equiv \langle \hat{m}_{\alpha} \rangle 
	\equiv \langle \hat{S}_{\alpha} \rangle / S$, where $\langle \hat{S}_\alpha \rangle \equiv  \operatorname{Tr} \left[ \hat{S}_\alpha \hat{\rho}(s) \right]$. The magnetization $m_{\alpha}$ serves as the order parameter of the BTC phase. The corresponding mean-field solution $m_{\alpha}^{\mathrm{MF}}$ approximates the magnetization and satisfies the following equations~\cite{PhysRevLett.121.035301,PhysRevA.110.012220,PhysRevA.105.L040202}
	\begin{equation*}
		\begin{aligned}
			\frac{\mathrm{d}}{\mathrm{d}t} m_x^{\mathrm{MF}} &=  \kappa m_x^{\mathrm{MF}} m_z^{\mathrm{MF}}, \\
			\frac{\mathrm{d}}{\mathrm{d}t} m_y^{\mathrm{MF}} &=  - \omega_0 m_z^{\mathrm{MF}} + \kappa m_y^{\mathrm{MF}} m_z^{\mathrm{MF}}, \\
			\frac{\mathrm{d}}{\mathrm{d}t} m_z^{\mathrm{MF}} &= \omega_0 m_y^{\mathrm{MF}} - \kappa  \left( m_x^{\mathrm{MF}} \right)^2 - \kappa \left( m_y^{\mathrm{MF}} \right)^2 .
		\end{aligned}
	\end{equation*}
	Here, the error between $m_{\alpha}^{\mathrm{MF}}$ and $m_{\alpha}$ is of order $1/N$. The dynamics is governed by a set of nonlinear differential equations, displaying non-decaying behavior in the long-time limit and approaching an asymptotic limit cycle rather than converging to a stationary fixed point (as indicated by the non-decaying oscillations in Fig.~\ref{fig1}a, b). 
	
	However, whether in numerical simulations or experimental realizations, the thermodynamic limit cannot be reached, and finite-size effects are unavoidable~\cite{yao2017discrete,gong2018discrete,article,liu2024higher,zhang2017observation,kongkhambut2022observation,kyprianidis2021observation,smits2018observation,rovny2018observation}. Consequently, the BTC relaxes to a steady state after long but finite time (see Fig.~\ref{fig1}e, f). As illustrated in Fig.~\ref{fig1}a, b, two conclusions can be drawn:  
	(i) the accuracy of the mean-field approximation improves with increasing particle number;  
	(ii) its validity deteriorates with increasing evolution time, becoming severe for asymptotically long times in finite-size systems.  
		   
	A deeper understanding of the above conclusions is provided by the theory developed in Refs.~\cite{carollo2024applicability,fiorelli2023mean}. 
	Let us introduce the error function $E_N(t)$:
	\begin{equation*}
		E_N(t) :=
		\sum_{\alpha}
		\left\langle
		\Phi(t)\big[ \left( \hat{m}_\alpha - m_\alpha^{\mathrm{MF}}
		\right)^2 \big]
		\right\rangle ,
	\end{equation*}
	where $\Phi(t)$ is the propagator of Eq.~\eqref{eq:1}. In the MFT, it satisfies the bound
	$E_N(t) \le C_2 e^{C_1 t}/N$,
	with positive constants $C_1$ and $C_2$.
	For a fixed time $t$, the inverse dependence of the error bound on the particle number indicates that the mean-field theory is valid and can be systematically improved order by order in the large-$N$ limit~\cite{PhysRevA.110.012220}.
%    A deeper understanding of the above conclusions is provided by the theory developed in Refs.~\cite{carollo2024applicability,fiorelli2023mean}. 
%    The error induced by the mean-field approximation is quantified by an error function $E_N(t)$, which is bounded as 
%    $E_N(t) \le C_2 e^{C_1 t}/N$, 
%    with positive constants $C_1$ and $C_2$. 
%    The error function is defined as 
%	\begin{equation*}
%		E_N(t) :=
%		\sum_{\alpha=1}^{(N+1)^2}
%		\left\langle
%		\Phi(t)\big[ \left( m_\alpha^{N} - m_\alpha(t)
%		\right)^2 \big]
%		\right\rangle,
%	\end{equation*}
%	where $m_\alpha(t)$ denotes the corresponding solution of the mean-field equations.
%	The propagator $\Phi(t)$ implements the exact many-body dynamics and is defined as the solution of the equation $\dot{\Phi}(t) = \Phi(t) \circ \mathcal{L}(t)$,
%	such that any operator $\hat{O}$ evolves according to 
%	$\hat{O}(t) = \Phi(t)[\hat{O}]$.
%	Giving a fixed time $t$, the error bound being inversely proportional to the particle number indicates that mean-field theory is valid and can be systematically improved order by order in the large $N$ limit~\cite{PhysRevA.110.012220}.
	 In contrast, when considering the full-time evolution from the initial state to the steady state in the BTC phase, the relaxation time scales as $N/\kappa$, which implies	 
	\begin{equation}
		E_{N}(t) < C_2 \frac{e^{C_1 N/\kappa}}{N}.
		\label{eq:3}
	\end{equation}
	The exponential growth of the error bound leads to the breakdown of the MFT. 
	To address these shortcomings of the standard mean-field treatment, which become severe in finite-size systems at asymptotically long times, we develop the SRWA. 
		
	Using the SRWA, the solution provides an accurate approximation to the order parameter obtained from numerical simulation (see Fig.~\ref{fig1}c, d). We also derive the analytical expressions for the steady-state density operator, the oscillation period and the decay rate of the order parameter. 
	The analytical expressions for the oscillation period $T$ and the decay rate $\gamma$ are given by
	\begin{align}
		\frac{T-T_0}{T_0}
		&=
		\frac{(4N^2+8N-11)\kappa^2}{8\omega_0^2N^2}
		\pm
		\frac{7\kappa^2 (N-1)}{2\omega_0^2 N^2}
		\, e^{-\frac{7\kappa t}{2N}},
		\label{eq:n1}
	\end{align}
	and
	\begin{equation}
		\gamma = -\frac{3\kappa}{2N},
		\label{eq:n2}
	\end{equation}
	where the initial state is chosen as the fully polarized spin-down state (see Sec.\ref{sec:results} for details).
	
	\section{stroboscopic rotating wave approximation}
	\label{sec:method}
	% 给出主要结果：公式、图示、对比分析
	%引用8、218、两篇引用
	In the finite-sized system, the evolution of the order parameter is characterized by two behaviors: decay and oscillation. The SRWA decomposes the dynamics into two dynamical regimes: the long-time evolution which captures the decay of the order parameter at stroboscopic points, and the short-time evolution which captures the oscillatory behavior between adjacent stroboscopic points. 
	
	We shift our discussions from the laboratory frame to the rotating frame~\cite{zeuch2020exact}, which is associated with a unitary transformation
	\begin{equation*}
		\hat{U}(t) = e^{-i \omega_{0} \hat{S}_{x} t}. 
	\end{equation*}
	Here, we conventionally take the rotation frequency to be $\omega_0$. In Sec.\ref{sec:new}, We will discuss the results obtained by adopting an alternative rotation frequency. 
	
	The dynamics of BTC in the rotating frame are given by (see Appendix~\ref{appendix:rotating} for more detials)
	\begin{align}
		\frac{\mathrm{d} \tilde{\rho}}{\mathrm{d} t} = \mathcal{L} \tilde{\rho} = \frac{\kappa}{N} \left(2 \tilde{S}_{-} \tilde{\rho} \tilde{S}_{+} - \tilde{S}_{+} \tilde{S}_{-} \tilde{\rho} - \tilde{\rho} \tilde{S}_{+} \tilde{S}_{-} \right),
		\label{eq:4}
	\end{align}
	where for any operator $\hat{O}(t)$, we define
	\begin{equation*}
		\tilde{O}(t) = \hat{U}^{\dagger}(t) \hat{O}(t) \hat{U}(t). 
	\end{equation*}
	
	In the high-frequency limit \(\omega_0 \to \infty\), we obtain the approximate master equation for the BTC with the help of RWA~\cite{zeuch2020exact},
	\begin{equation}
		\begin{aligned}
			\frac{\mathrm{d} \tilde{\rho}}{\mathrm{d} t} & = \frac{\kappa}{N} \bigg( 
			2\hat{S}_x \tilde{\rho} \hat{S}_x -  \hat{S}_x ^2 \tilde{\rho} -  \tilde{\rho} \hat{S}_x ^2 \\
			& \quad +  \hat{S}_y \tilde{\rho} \hat{S}_y - \frac{1}{2} \hat{S}_y ^2 \tilde{\rho} - \frac{1}{2} \tilde{\rho} \hat{S}_y ^2 \\
			& \quad + \hat{S}_z \tilde{\rho} \hat{S}_z - \frac{1}{2} \hat{S}_z ^2 \tilde{\rho} - \frac{1}{2} \tilde{\rho} \hat{S}_z ^2 
			\bigg).
		\end{aligned}
		\label{eq:5}
	\end{equation}
	In Eq.~\eqref{eq:5}, the three terms on the right-hand side represent dephasing processes along the $X$, $Y$, and $Z$ directions, respectively. 
	The competition among dephasing along three directions induces persistent oscillations of the order parameter, signaling the formation of the BTC phase. 	
	
	For a finite $\omega_{0}$, we perform higher-order approximation using the SRWA.  
	The system evolves from time $0$ to $t$, and we decompose $t$ as  
	\begin{equation}
		t = r_{n} + s,
		\label{eq:6}
	\end{equation}
	where $r_{n} = nT_{0}$, $n$ is a natural number,  $s \in [0, T_{0})$ is the remainder, and the times $\{0, T_{0}, 2T_{0}, \dots\}$ are called stroboscopic points.
	Then, we decompose the solution of Eq~\eqref{eq:5}, i.e., $\tilde{\rho}(t) =\mathcal{V}(t)\hat{\rho}(0) $ as
	\begin{equation}
		\mathcal{V}(t) = \mathcal{V}(s)\,\mathcal{V}(r_n),
		\label{eq:7}
	\end{equation}
	where $\mathcal{V}(t) = \mathcal{T} 
	e^{ \int_{0}^{t} \mathrm{d}\tau\, \mathcal{L}(\tau) }$ denotes evolution superoperator of Eq~\eqref{eq:5}, and $ \mathcal{T} $ is the time ordering operator. Here $\mathcal{V}(r_n)$ represents the long-time evolution at the stroboscopic points, and 
	$\mathcal{V}( s)$ corresponds to the short-time evolution between adjacent stroboscopic 
	points.
	
	We furthermore simplify the long-time evolution by using the Magnus expansion~\cite{zeuch2020exact,blanes2009magnus}:
	\begin{equation}
		\mathcal{V}(r_n) = \mathcal{T} 
		e^{ \int_{0}^{t} \mathrm{d}\tau\, \mathcal{L}(\tau) }
		= e^{r_n  \bar{\mathcal{L}} },
		\label{eq:8}
	\end{equation}	
	where the effective Lindblad superoperator  $ \bar{\mathcal{L}} $ is given by
	\begin{equation}
		\bar{\mathcal{L}} = \sum_{m=0}^{\infty} \bar{\mathcal{L}}^{(m)}
		\label{eq:9}
	\end{equation}	
	with the first several terms given by:
	\begin{align}
		\bar{\mathcal{L}}^{(0)} &= \frac{1}{r_n} \int_{0}^{r_n} \mathrm{d}\tau\, \mathcal{L}(\tau), \label{eq:10a} \\[6pt]
		\bar{\mathcal{L}}^{(1)} &= \frac{1}{2 r_n} \int_{0}^{r_n} \mathrm{d}\tau' \int_{0}^{\tau'} \mathrm{d}\tau\, 
		\left[ \mathcal{L}(\tau'), \mathcal{L}(\tau) \right], \label{eq:10b}\\[6pt]
		\bar{\mathcal{L}}^{(2)} &= \frac{1}{6 r_n} \int_{0}^{r_n} \mathrm{d}\tau'' \int_{0}^{\tau''} \mathrm{d}\tau' \int_{0}^{\tau'} \mathrm{d}\tau\,
		\bigg\{ \left[ \mathcal{L}(\tau''), \left[ \mathcal{L}(\tau'), \mathcal{L}(\tau) \right] \right] \notag \\[-3pt]
		&\quad
		+ \left[ \left[ \mathcal{L}(\tau''), \mathcal{L}(\tau') \right], \mathcal{L}(\tau) \right] \bigg\} \label{eq:10c}.
	\end{align}	
	By truncating at the first order of $\kappa/\omega_{0}$, $\bar{\mathcal{L}} \simeq \bar{\mathcal{L}}^{(0)} + \bar{\mathcal{L}}^{(1)}$, we obtain the expressions of  $\bar{\mathcal{L}}^{(0)}$ (Eq.~\ref{eq:5}) and $\bar{\mathcal{L}}^{(1)}$ (Eq.~\ref{eq:11} in Appendix~\ref{appendix:longtime and shorttime}).

	Also, we approximate the short-time evolution using a Taylor expansion:
	\begin{align}
		\mathcal{V}(s) &= \mathcal{T} e^{\int_0^{s} \mathrm{d}\tau\, \mathcal{L}(\tau)} \notag \\
		& =\sum_{m=0 }^{\infty} \frac{1}{m!} \mathcal{G}(s) ^m \notag \\
		& = \mathbbm{1} + \mathcal{G}(s) + \mathcal{O}\left(\frac{\kappa^2}{\omega^2_0}\right),
		\label{eq:12}
	\end{align}  
	where $ \mathcal{G}(s) \equiv  \int_0^s d\tau\, \mathcal{L}(\tau)$, and $\mathbbm{1}$ is the identity superoperator. By truncating at the first order of $\kappa/\omega_{0}$, we obtain the expression of the reduced quantum dynamical semigroup $ \mathcal{G}(s)$ \cite{footnote1,HIGHAM1988103} (Eq.~\ref{eq:13} in Appendix~\ref{appendix:longtime and shorttime}).

	\section{Results}
	\label{sec:results}
	% 给出主要结果：公式、图示、对比分析
	Building upon the SRWA developed in Sec.~\ref{sec:method}, we study the full-time dynamics of BTC in finite-size systems and obtain the analytical expressions for the steady-state density operator, the oscillation period and the decay rate of the order parameter, which go beyond the results obtained by using the MFT.	 
	
	\begin{figure}[t]
		\renewcommand{\figurename}{FIG.}
		\centering
		
		% --- Row 1 ---
		\includegraphics[scale=0.18]{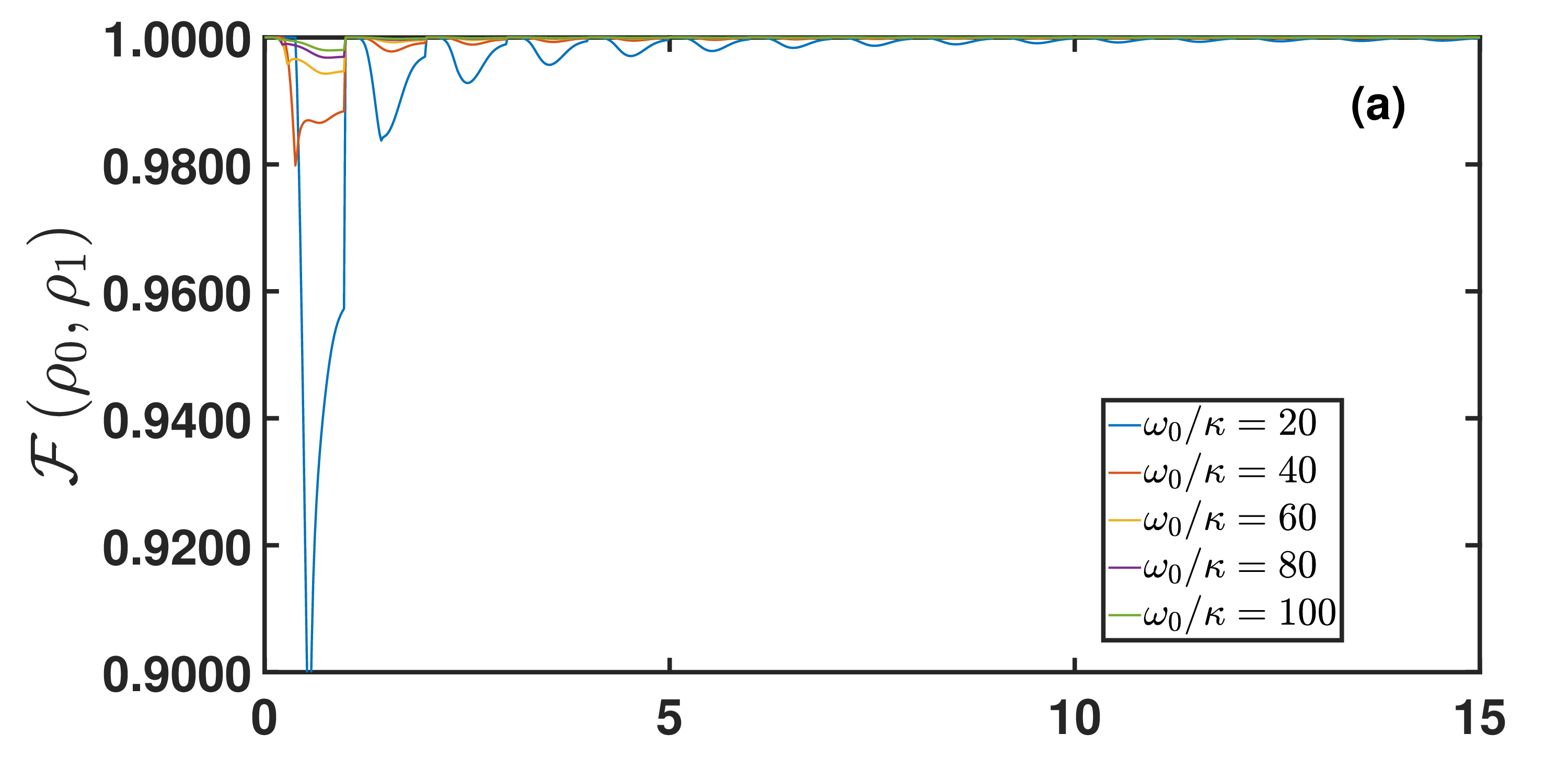}
		
		\vspace{6pt}
		
		% --- Row 2 ---
		\includegraphics[scale=0.18]{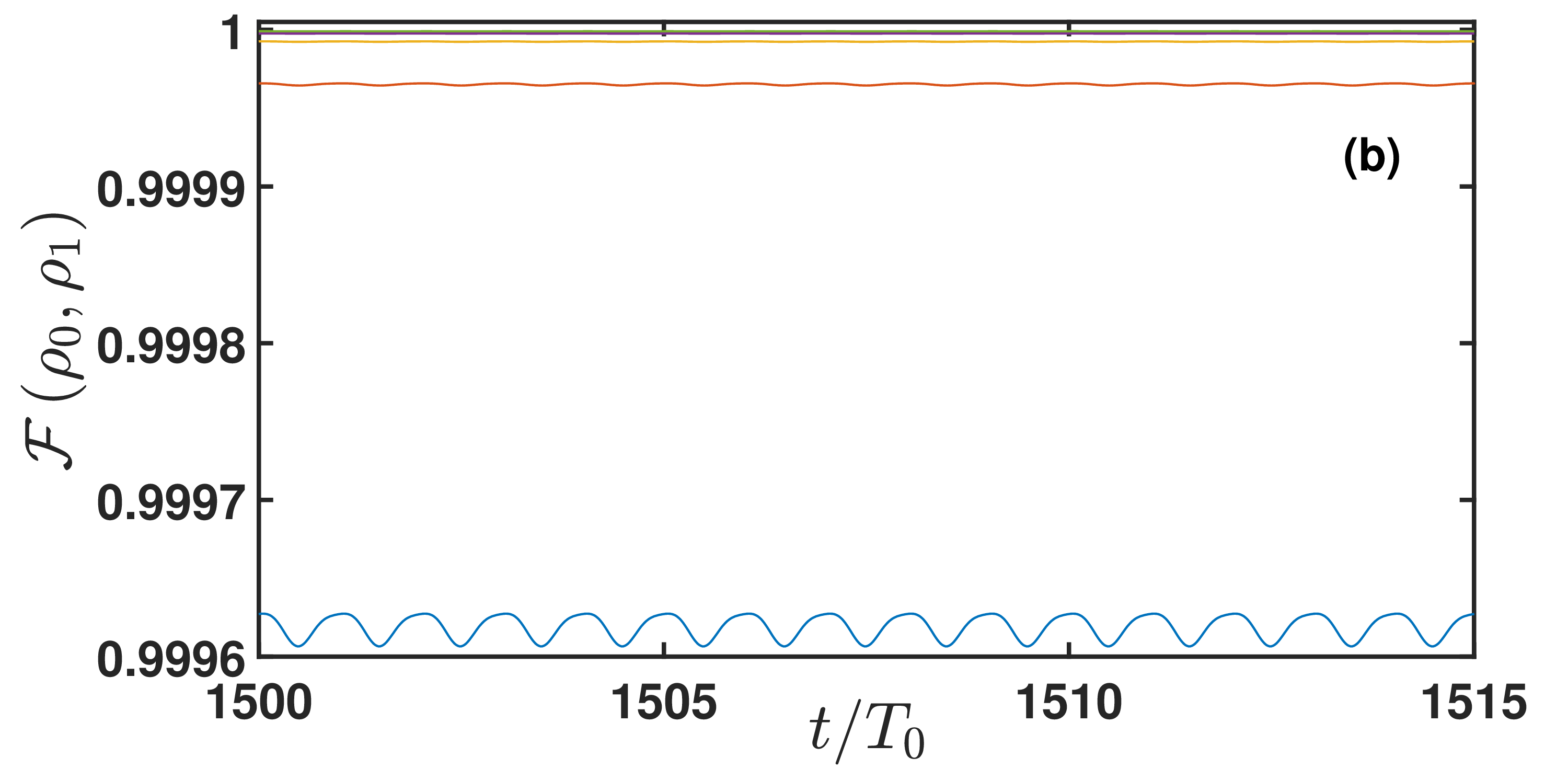}
		
		\vspace{6pt}

		\caption{Time evolution of the fidelity between the density matrices obtained by using the numerical simulation of Eq.~\eqref{eq:1} and the SRWA ($N=50$, and the initial state is all spins pointing down). Panels (a) and (b) show the fidelity evaluated over the driving-period intervals $0$--$15$ and $1500$--$1515$, respectively.}
		\label{fig4}
	\end{figure}
	
	\subsection{Construction of full-time dynamics by using the SRWA}
	
	Returning to the laboratory frame from the rotating frame, we approximate the full-time evolution of the BTCs Eqs.~\eqref{eq:7}-\eqref{eq:12}:
	\begin{equation}
		\hat{\rho}(t) = \hat{U}(t) \left \{  \left[ \mathbbm{1} + \mathcal{G}(s) \right] e^{r_n(\bar{\mathcal{L}}^{(0)} + \bar{\mathcal{L}}^{(1)})} \hat{\rho}(0) \right \}  \hat{U}^{\dagger}(t) + \mathcal{O}\!\left(\frac{\kappa^2}{\omega^2_0}\right).
		\label{eq:14}
	\end{equation}
	
	To demonstrate the accuracy of the SRWA, We adopt the following definition of Uhlmann fidelity between the density matrices $\rho_0$ and $\rho_1$ obtained by using the numerical simulation of Eq.~\eqref{eq:1} and the SRWA, respectively~\cite{jozsa1994fidelity,ma2009many}.
	\begin{equation*}
		\mathcal{F}(\rho_0, \rho_1)
		= \left[ \operatorname{Tr} \left( 
		\sqrt{ \sqrt{\rho_0} \, \rho_1 \, \sqrt{\rho_0} } 
		\right) \right]^2 .
	\end{equation*} 
	As shown in Fig.~\ref{fig4}, we have:  
	(i) For a fixed time $t$, the accuracy of the SRWA results increases with the driving frequency~$\omega_0$;  
	(ii) For a fixed $\omega_0$, the accuracy of the SRWA increases as time increases, and it saturates at long times when the system reaches the steady state ~\cite{footnote2}. 
		
	\subsection{Steady-state in the BTC phase}
	
	The steady-state of the BTC phase is the result of the long-time evolution, which satisfies:
	\begin{equation}
		\bar{\mathcal{L}} \hat{\rho}_s =0 .
		\label{eq:15}
	\end{equation}
    We obtain an approximate solution for the steady-state density matrices $\hat{\rho}_s$ using perturbation theory. Let $\bar{\mathcal{L}} = \bar{\mathcal{L}}^{(0)} + \bar{\mathcal{L}}^{(1)}$, $\hat{\rho}_s = \hat{\rho}_s^{(0)} + \hat{\rho}_s^{(1)}$, and substitute them into Eq.~\eqref{eq:15}. Then we have
	\begin{align}
		\bar{\mathcal{L}}^{(0)} \hat{\rho}_s^{(0)} &= 0, \label{eq:16}\\
		\bar{\mathcal{L}}^{(0)} \hat{\rho}_s^{(1)} &= - \bar{\mathcal{L}}^{(1)} \hat{\rho}_s^{(0)}. \label{eq:17}
	\end{align}
	According to Eq.~\eqref{eq:5}, $\bar{\mathcal{L}}^{(0)}$ induces dephasing along the $X$, $Y$, and $Z$ directions, implying that $\hat{\rho}_s^{(0)}$ must be a maximally mixed state. Similar steady-state structures have been reported in cooperative resonance fluorescence~\cite{carmichael1980analytical,drummond1980observables}.
	\begin{equation}
		\hat{\rho}_s^{(0)} = \frac{1}{N+1} \mathbbm{1}.
		\label{eq:18}
	\end{equation}
	Substituting Eq.~\eqref{eq:18} into Eq.~\eqref{eq:17}, we obtain the first-order correction to the steady-state density as
	\begin{equation}
		\hat{\rho}_s^{(1)} = \frac{4\kappa}{\omega_0 N} \hat{S}_y.
		\label{eq:19}
	\end{equation}
	We find that Eqs.~\eqref{eq:18} and \eqref{eq:19} coincide with the exact solution given in Ref.~\cite{drummond1980observables} in the limit $\omega_0 \to \infty$.
	\begin{figure}[t]
		\renewcommand{\figurename}{FIG.}
		\centering 
		\includegraphics[scale=0.25]{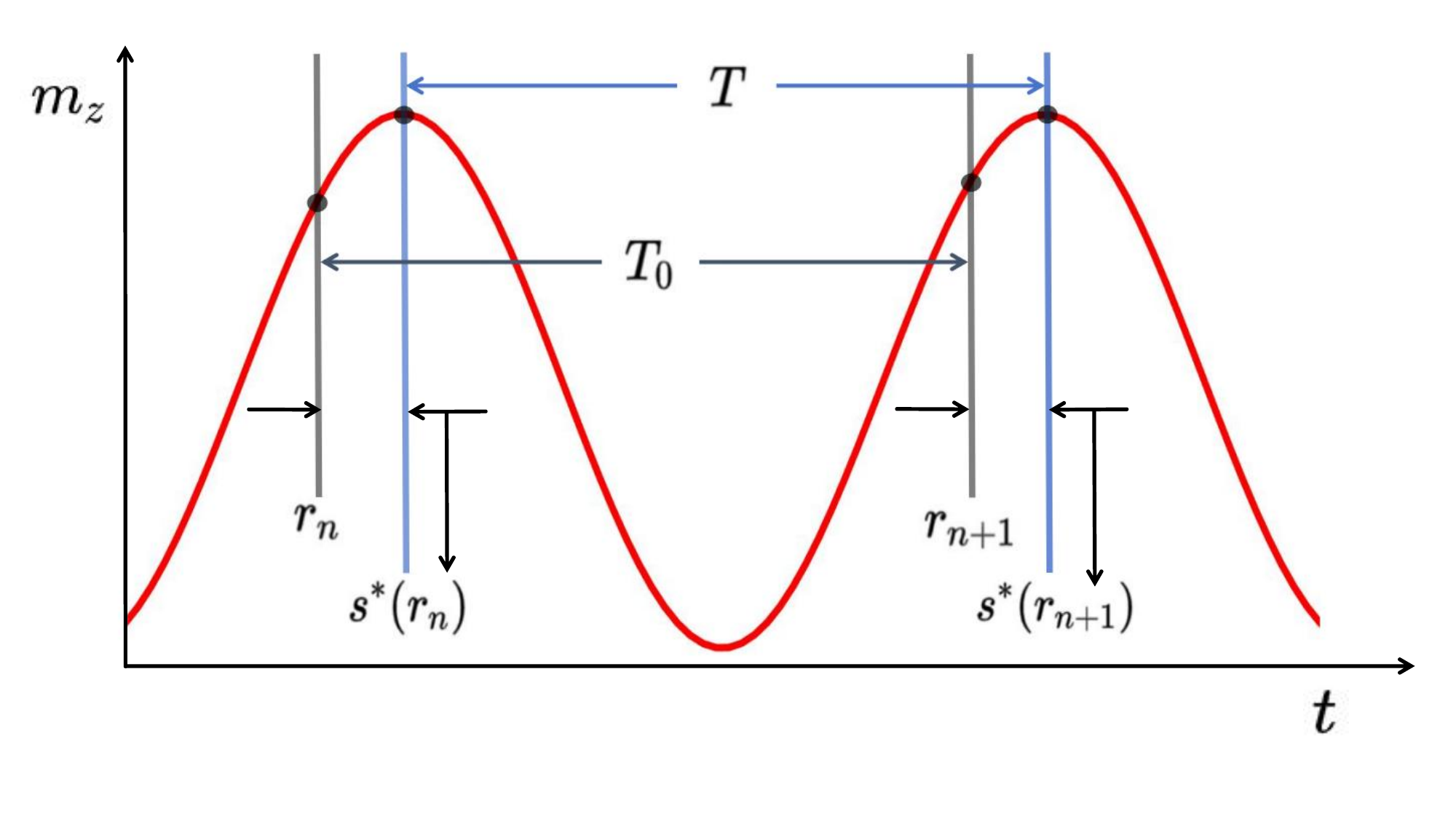}
		\caption{Schematic diagram for calculating the oscillation period of the order parameter $m_z$. The red curve represents the time evolution of $m_z$. $r_n$ and $r_{n+1}$ denote two adjacent stroboscopic points. $s^{*}(r_n)$ and $s^{*}(r_{n+1})$ denote the times of the local maximum points, whose values correspond to the time interval between the local maximum points and the stroboscopic points.
		$T_0$ and $T$ represent the driving period and the oscillation period, respectively.}
		\label{fig5}  
	\end{figure}
	
	\subsection{\texorpdfstring{Oscillation period of $\langle \hat{S}_z \rangle$}{Oscillation period of Sz}}
	
	As shown in Fig.~\ref{fig1}, $\langle \hat{S}_{z} \rangle$ exhibits a damped oscillation over time. 
	Using Eqs.~\eqref{eq:12} and~\eqref{eq:14}, we obtain an analytical expression for $\langle \hat{S}_z \rangle$:	
	\begin{align}
		&\langle \hat{S}_{z} \rangle (r_n, s) \notag \\
		& = \left(1- \frac{3 \kappa s }{2 N}\right)
		\Big[ \cos(\omega_0 s)\langle \hat{S}_z \rangle (r_n) 
		+ \sin(\omega_0 s)\langle \hat{S}_y \rangle (r_n) \Big] \notag\\
		& + \frac{\kappa}{2 \omega_0 N} \Big[
		2 \sin(2\omega_0 s)
		\langle \hat{S}_z^2 - \hat{S}_y^2 \rangle (r_n) \notag \\
		& + 2 \cos(\omega_0 s) 
		\langle \hat{S}_y \hat{S}_z + \hat{S}_z \hat{S}_y \rangle (r_n) \notag \\
		& - \sin(\omega_0 s) \langle \hat{S}_z \rangle (r_n)  
		- 4 \sin(\omega_0 s) 
		\langle \hat{S}_x^2 + \hat{S}_z^2 \rangle (r_n) \notag \\
		& - 2 \cos(2\omega_0 s) 
		\langle \hat{S}_y \hat{S}_z + \hat{S}_z \hat{S}_y \rangle (r_n) \Big].
		\label{eq:21}
	\end{align}
%	Here, due to the time decomposition in Eq.~\eqref{eq:6}, $\langle \hat{O} \rangle (r_n, s) = \langle \hat{O} \rangle(t)$ is considered as a bivariate function of $r_n$ and $s$, and $\langle \hat{O} \rangle (r_n)$ represents the expectation value for operator $\hat{O}$ at the stroboscopic point $r_n$. The time interval between two adjacent stroboscopic points, $r_n$ and $r_{n+1}$, corresponds to the driving period $T_0$ rather than the oscillation period $T$ of $\langle \hat{S}_z \rangle$. 
    The expectation value $\langle \hat{O} \rangle (r_n, s) = \langle \hat{O} \rangle(t)$ is treated as a function of $r_n$ and $s$, with $\langle \hat{O} \rangle(r_n)$ denoting its value at the stroboscopic point $r_n$. The time interval between two adjacent stroboscopic points, $r_n$ and $r_{n+1}$, corresponds to the driving period $T_0$, which differs from the oscillation period $T$ of $\langle \hat{S}_z \rangle$.
	As shown in Fig.~\ref{fig5}, we define the oscillation period $T$ as the time interval between its two adjacent local maximum or minimum points, which satisfy
	\begin{equation}
		\frac{\partial \langle \hat{S}_{z} \rangle  }{\partial s}\bigg|_{s=s^*} = 0.
		\label{eq:22}
	\end{equation}
	By substituting Eq.~\eqref{eq:21} into Eq.~\eqref{eq:22}, we obtain
	\begin{equation}
		\tan(\omega_0 s^*) = f(r_n, s^*),
		\label{eq:23}
	\end{equation}
%	\begin{widetext}
%		\begin{align}
%			f(r_n, s^*) &= 
%			\frac{\langle \hat{S}_y \rangle}{\langle \hat{S}_z \rangle}  		
%			- \frac{\kappa}{2 \omega_0 N \langle \hat{S}_z \rangle^2}
%			\Bigg[  
%			3 \langle \hat{S}_y \rangle^2 
%			+ 2 \langle \hat{S}_y \rangle \langle \hat{S}_y \hat{S}_z + \hat{S}_z \hat{S}_y \rangle 
%			+ 4 \langle \hat{S}_z \rangle \bigl( \langle \hat{S}_x^2 + \hat{S}_z^2 \rangle + \langle \hat{S}_z \rangle \bigr) \notag\\
%			&\quad 
%			+ 4 \langle \hat{S}_z \rangle \bigl( \langle \hat{S}_y^2 - \hat{S}_z^2 \rangle \bigr) \cos(\omega_0 s^*) \sec(\omega_0 s^*) 
%			- 8 \langle \hat{S}_y \rangle \langle \hat{S}_y \hat{S}_z + \hat{S}_z \hat{S}_y \rangle \langle \hat{S}_z \rangle 
%			\Bigg] 
%			+ \mathcal{O}\!\left(\frac{\kappa^2}{\omega^2_0}\right).
%			\label{eq:24}
%		\end{align}
%	\end{widetext}
	where
    \begin{align}
    	f(r_n, s^*) &= 
    	\frac{\langle \hat{S}_y \rangle}{\langle \hat{S}_z \rangle}  
    	- \frac{\kappa}{2 \omega_0 N \langle \hat{S}_z \rangle^2} \Bigg[
    	 2 \langle \hat{S}_y \rangle 
    	\langle \hat{S}_y \hat{S}_z + \hat{S}_z \hat{S}_y \rangle 
    	\notag\\
    	&\quad 
    	+ 4 \langle \hat{S}_z \rangle 
    	\langle \hat{S}_x^2 + \hat{S}_z^2 \rangle
    	+ 4 \langle \hat{S}_z \rangle 
    	\langle \hat{S}_y^2 - \hat{S}_z^2 \rangle
    	 \notag\\
    	&\quad
    	- 8 \langle \hat{S}_y \rangle 
    	\langle \hat{S}_y \hat{S}_z + \hat{S}_z \hat{S}_y \rangle 
    	\langle \hat{S}_z \rangle \notag\\
    	&\quad
    	+ 3 \langle \hat{S}_y \rangle^2 
    	+ 4 \langle \hat{S}_z \rangle^2
    	\Bigg] 
    	+ \mathcal{O}\!\left(\frac{\kappa^2}{\omega_0^2}\right) .
    	\label{eq:24}
    \end{align}	  			 
	The time dependence of the  expectation values on $r_n$ is omitted for simplicity.
	
	The oscillation period in Eq.~\eqref{eq:24} is related to the driving period as follows (see Fig.~\ref{fig5}):
	\begin{equation}
		T = T_0 + s^*(r_{n+1}) - s^*(r_n), 
		\label{eq:25}
	\end{equation}	
	The extremum point $s^*$ varies slowly with $r_n$. By performing Taylor expansion, 
	we approximate Eq.~\eqref{eq:25} as
%	we have $s^*(r_{n+1}) \simeq s^*(r_n) + T_0 \frac{\mathrm{d} s^*(r_n)}{\mathrm{d} r_n}$, and	
	\begin{equation}	
		\frac{T - T_0}{T_0} = \frac{\mathrm{d} s^*(r_n)}{\mathrm{d} r_n}. \label{eq:26}
	\end{equation}
	Differentiating both side of Eq.~\eqref{eq:23} with respect to $r_n$, we obtain
	\begin{equation}
		\frac{\mathrm{d} s^*}{\mathrm{d} r_n} 
		= \frac{1}{\omega_0 \left[ 1 + \tan^2(\omega_0 s^*) \right] } \left( \frac{\partial f}{\partial r_n} + \frac{\partial f}{\partial s^*} \frac{\mathrm{d} s^*}{\mathrm{d} r_n} \right),
		\label{eq:27}
	\end{equation}
	where the partial derivative is taken with respect to $r_n$ while keeping $s^*$ fixed. 
	By substituting Eqs.~\eqref{eq:26} into~\eqref{eq:27} and neglecting the second term on the right-hand side of Eq.~\eqref{eq:27} (which is of second order in $\kappa/\omega_{0}$), we obtain:
	\begin{equation}
		\frac{T - T_0}{T_0} 
		= \frac{1}{\omega_0 \left[ 1 + \tan^2(\omega_0 s^*) \right]}  \frac{\partial f}{\partial r_n} + \mathcal{O}\left(\frac{\kappa^2}{\omega^2_0}\right).
		\label{eq:28}
	\end{equation}
	
\begin{figure}[t]
		\renewcommand{\figurename}{FIG.}
		\centering 
		\includegraphics[scale=0.3]{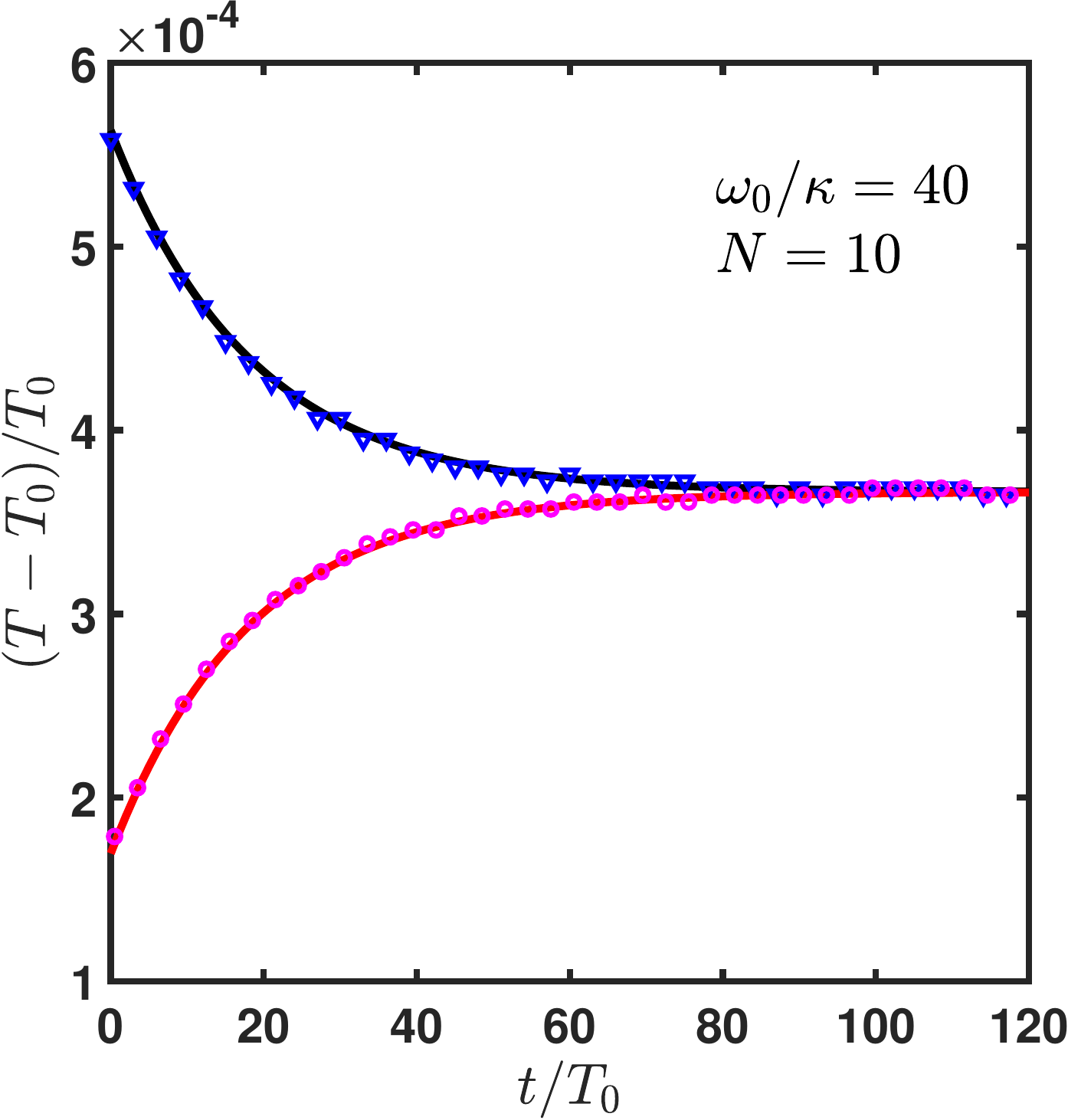}
		\caption{Time dependence of the relative deviation between the oscillation period and the driving period of $\langle \hat{S}_z \rangle$.
			The initial state is chosen with all spins pointing down.
			The blue inverted triangle and purple circles represent the relative deviations calculated from the local maximum or minimum points of $\langle \hat{S}_{z} \rangle$ from the numerical simulation of Eq.~\eqref{eq:1}, respectively. 
			The black and red solid lines correspond to the results from Eq.~\eqref{eq:30}.}
		\label{fig2}  
	\end{figure}

	Next, we use Eqs.~\eqref{eq:23} and~\eqref{eq:24} to simplify the trigonometric functions in Eq.~\eqref{eq:28}. The first term on the right-hand side of Eq.~\eqref{eq:24} is the leading-order. Then the solution for $s^*$ in Eq.~\eqref{eq:23} is
	\begin{equation}
		s^* = \frac{1}{\omega_0} \tan\!\left( \frac{\langle \hat{S}_y \rangle}{\langle \hat{S}_z \rangle} \right)  + \mathcal{O}\left(\frac{\kappa}{\omega^2_0}\right).
		\label{eq:29}
	\end{equation}
	By substituting Eq.~\eqref{eq:29} into Eq.~\eqref{eq:24} and using Eq.~\eqref{eq:28}, we obtain 
	\begin{align}
		\frac{T - T_0}{T_0} 
		&= \frac{(4 N^2 + 8 N - 11)\kappa^2}{8 \omega_0^2 N^2} 
		\notag\\
		&\quad \pm \frac{7 \kappa^2}{\omega_0^2 N^2 (\langle \hat{S}_y \rangle^2 + \langle \hat{S}_z \rangle^2)^{3/2}} 
		\Big[
		 \langle \hat{S}_y^2 - \hat{S}_z^2 \rangle \langle \hat{S}_y \rangle^2
		\notag\\
		&\quad + 2 \langle \hat{S}_y \rangle \langle \hat{S}_z \rangle 
		\langle \hat{S}_y \hat{S}_z + \hat{S}_z \hat{S}_y \rangle \notag\\
		&\quad + \langle \hat{S}_z^2 - \hat{S}_y^2 \rangle \langle \hat{S}_z \rangle^2
		\Big]
		+ \mathcal{O}\left(\frac{\kappa^2}{\omega^2_0}\right).
		\label{eq:30}
	\end{align}   
%	\begin{widetext}
%		\begin{align}
%			\frac{T - T_0}{T_0} 
%			&= \frac{(4 N^2 + 8 N - 11)\kappa ^{2}}{8\omega ^{2}_{0}N^{2}}  
%			\pm \frac{7\kappa ^{2}}{\omega ^{2}_{0}N^{2}}  
%			\frac{
%				2 \langle \hat{S}_y  \rangle  \langle \hat{S}_z  \rangle  \langle \hat{S}_y \hat{S}_z + \hat{S}_z \hat{S}_y  \rangle
%				+  \langle \hat{S}_y^2  - \hat{S}_z^2  \rangle \langle \hat{S}_y  \rangle^2
%				+  \langle \hat{S}_z^2 - \hat{S}_y^2  \rangle  \langle \hat{S}_z  \rangle^2
%			}
%			{\left( \langle \hat{S}_y  \rangle^2 +  \langle \hat{S}_z  \rangle^2 \right)^{3/2}}
%			+ \mathcal{O}\!\left(\frac{\kappa^2}{\omega^2_0}\right).
%			\label{eq:30}
%		\end{align}
%	\end{widetext}	
	The $\pm$ sign corresponds to the local maximum or minimum points of $\langle \hat{S}_{z} \rangle$, respectively. 
	Substituting the long-time $r_n$-dependent solutions 
	[Eqs.~\eqref{eq:c8}--\eqref{eq:c14}] into Eq.~\eqref{eq:30}, 
	we obtain (see Appendix~\ref{appendix:SRWA})	
	\begin{align}
		\frac{T - T_0}{T_0} 
		&= \frac{(4 N^2 + 8 N - 11)\kappa^2}{8 \omega_0^2 N^2} 
		\notag\\
		&\quad \pm \frac{7 \kappa^2 e^{-\frac{7\kappa r_n}{2 N}}}{\omega_0^2 N^2 (\langle \hat{S}_y \rangle_0^2 + \langle \hat{S}_z \rangle_0^2)^{3/2}} 
		\Big[
		 \langle \hat{S}_y^2 - \hat{S}_z^2 \rangle_0 \langle \hat{S}_y \rangle_0^2
		\notag\\
		&\quad
		+ 2 \langle \hat{S}_y \rangle_0 \langle \hat{S}_z \rangle_0 
		\langle \hat{S}_y \hat{S}_z + \hat{S}_z \hat{S}_y \rangle_0
		\notag\\
		&\quad
		+ \langle \hat{S}_z^2 - \hat{S}_y^2 \rangle_0 \langle \hat{S}_z \rangle_0^2
		\Big]
		+ \mathcal{O}\left(\frac{\kappa^2}{\omega^2_0}\right).
		\label{eq:31}
	\end{align} 
%	\begin{widetext}
%		\begin{align}
%			\frac{T - T_0}{T_0} & = \frac{(4 N^2 + 8 N - 11)\kappa ^{2}}{8\omega ^{2}_{0}N^{2}} \pm  \frac{7\kappa ^{2}}{\omega ^{2}_{0}N^{2}}  \frac{
%				2 \langle \hat{S}_y  \rangle_{0}  \langle \hat{S}_z  \rangle_{0}  \langle \hat{S}_y \hat{S}_z + \hat{S}_z \hat{S}_y \rangle_{0}
%				+  \langle \hat{S}_y^2  -  \hat{S}_z^2  \rangle_{0}   \langle \hat{S}_y \rangle_{0}^2
%				+  \langle \hat{S}_z^2 - \hat{S}_y^2  \rangle_{0}   \langle \hat{S}_z  \rangle_{0}^2
%			}
%			{\left(  \langle \hat{S}_y  \rangle_{0}^2 +  \langle \hat{S}_z  \rangle_{0}^2 \right)^{3/2}} e^{-\frac{7\kappa r_n}{2 N}} + \mathcal{O}\left(\frac{\kappa^2}{\omega^2_0}\right),
%			\label{eq:31}
%		\end{align}
%	\end{widetext}	
	where $\langle \hat{O} \rangle_{0}$ represents the expectation value of operator $\hat{O}$ in the initial state. The first term on the right-hand side represents the correction to the oscillation period near the steady state. 
	In the limit $N \to \infty$,
	\begin{align}
		 \frac{T-T_0}{T_0} = \frac{\kappa^2}{2 \omega _0 ^2},
		 \label{eq:101}
	\end{align}
	which is consistent with the result $\omega = \sqrt{\omega_0^2 - \kappa^2}$ obtained by using the MFT in the limit $\omega_0 \to \infty$~\cite{PhysRevA.105.L040202}. 
	The second term describes the exponentially decaying memory effect. 
	The curves in Fig.~\ref{fig2} demonstrate that our solution (Eq.~\ref{eq:31}) closely fits the simulated results of Eq.~\eqref{eq:1}.
				
	\subsection{\texorpdfstring{Decay rate of $\langle \hat{S}_z \rangle$}{Oscillation period of Sz} }
	\label{sec:decay}
	
	\begin{figure}[t]
		\renewcommand{\figurename}{FIG.}
		\centering 
		\includegraphics[scale=0.3]{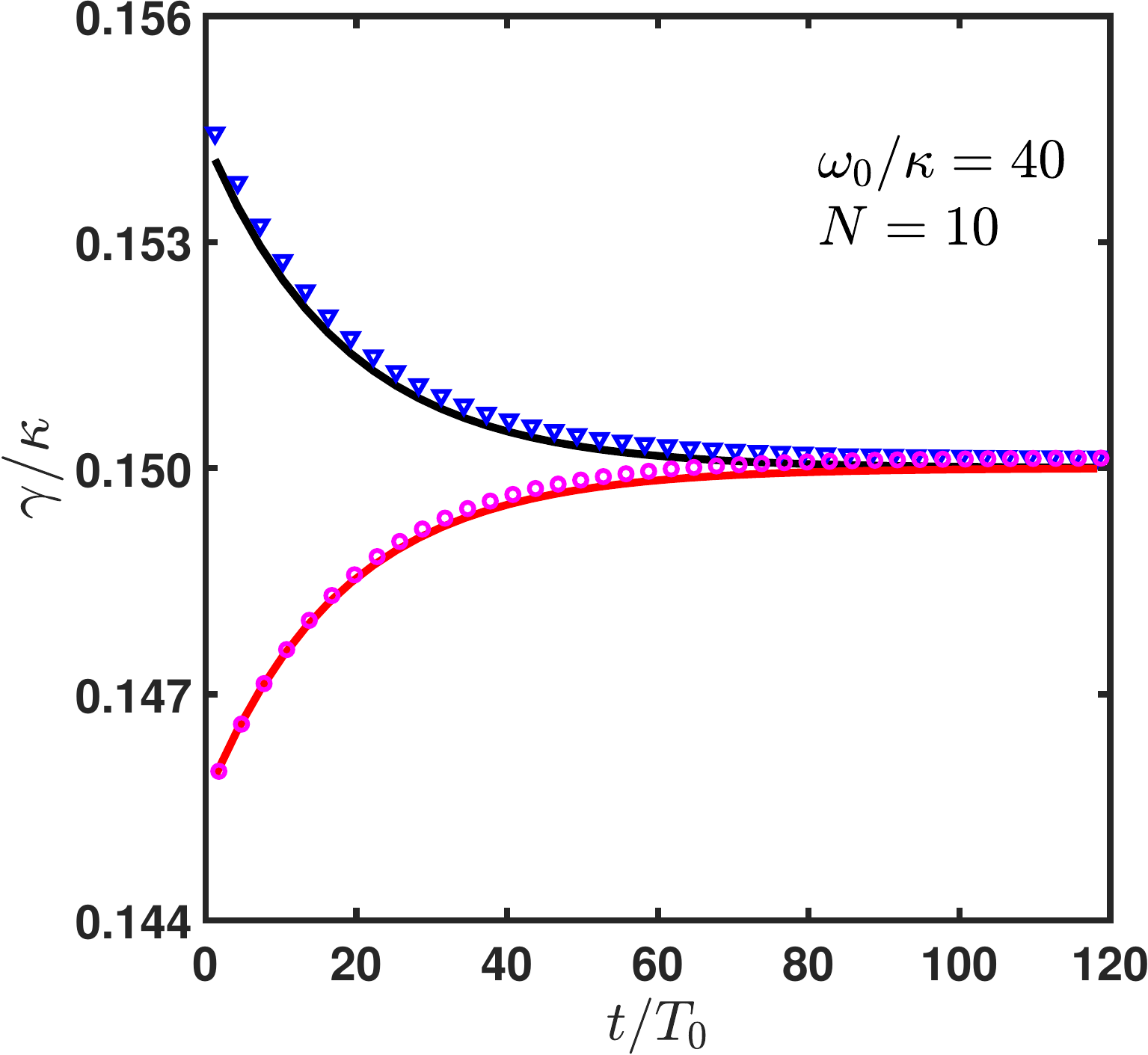}
		\caption{Time dependence of the decay rate of $\langle \hat{S}_z \rangle$.
			The initial state is chosen with $e^{-0.2i\left ( \hat{S}_y - \hat{S}_z  \right )^2 } \left | \frac{\pi}{3} ,\frac{\pi}{2}   \right \rangle $.
			The blue inverted triangle and purple circles indicate $\gamma$ calculated from the local maximum or minimum points of $\langle \hat{S}_{z} \rangle$ from the numerical simulation of Eq.~\eqref{eq:1}, respectively. The black and red solid lines correspond to the results from Eq.~\eqref{eq:34}.}
		\label{fig3}  
	\end{figure}
	
	We define the decay rate $\gamma$ as the rate of change of $\langle \hat{S}_z \rangle$ relative to itself, i.e.,
	\begin{equation}
		\gamma \equiv - \frac{1}{\langle \hat{S}_z \rangle}\frac{\mathrm{d}}{\mathrm{d}t}\langle \hat{S}_z \rangle .
		\label{eq:32}
	\end{equation}
    By using the expressions of $ \bar{\mathcal{L}}^{(0)} $ and $ \bar{\mathcal{L}}^{(1)}$ in Eqs.~\eqref{eq:b4} and ~\eqref{eq:11}, we obtain the long-time evolution of $\langle \hat{S}_{z} \rangle$ at the stroboscopic points, 	  
	\begin{align}
		\frac{\partial \langle \hat{S}_{z}  \rangle  }{\partial r_n}  
		&= -\frac{3\kappa }{2 N} \langle \hat{S}_{z}  \rangle 
		- \frac{ \kappa^{2}}{2 \omega N^{2}} 
		\Bigg[ \left( N^2 + 2N - \frac{11}{4} \right)\langle \hat{S}_{y}  \rangle \notag \\[1mm]
		&\quad - 7 \langle  \hat{S}_{y}\hat{S}_{z} + \hat{S}_{z}\hat{S}_{y} \rangle    \Bigg] + \mathcal{O}\left(\frac{\kappa^2}{\omega^2_0}\right) .
		\label{eq:33}
	\end{align}
	Then, by substituting the $r_n$-dependent long-time solutions for the expectation values, Eqs.~\eqref{eq:c8}--\eqref{eq:c14}, into Eq.~\eqref{eq:32} (see Appendix~\ref{appendix:SRWA}) and performing a Taylor expansion, we obtain:
	\begin{align}
		\gamma 
		&= -\frac{3\kappa}{2 N} 
		\notag\\
		&\quad \pm \frac{7\kappa^2 e^{-\frac{7 \kappa r_n}{2 N}}}{2\omega_0 N^2 (\langle \hat{S}_z \rangle_0^2 + \langle \hat{S}_y \rangle_0^2)^{3/2}} 
		\Big[
		2 \langle \hat{S}_y \rangle_0 \langle \hat{S}_z \rangle_0 \langle \hat{S}_y^2 - \hat{S}_z^2 \rangle_0
		\notag\\
		&\quad
		+ (\langle \hat{S}_z \rangle_0^2 - \langle \hat{S}_y \rangle_0^2) \langle \hat{S}_y \hat{S}_z + \hat{S}_z \hat{S}_y \rangle_0
		\Big]
		+ \mathcal{O}\left(\frac{\kappa^2}{\omega^2_0}\right).
		\label{eq:34}
	\end{align}	  
%	\begin{widetext}
%		\begin{equation}
%			\gamma = -\frac{3\kappa}{2 N} 
%			\pm \frac{7\kappa^2}{2\omega_0 N^2}   \frac{(\langle \hat{S}_z \rangle_0^2 - \langle \hat{S}_y \rangle_0^2) \langle \hat{S}_y \hat{S}_z + \hat{S}_z \hat{S}_y \rangle_0 
%				+ 2 \langle \hat{S}_y \rangle_0 \langle \hat{S}_z \rangle_0 \langle \hat{S}_y^2 - \hat{S}_z^2 \rangle_0}
%			{(\langle \hat{S}_z \rangle_0^2 + \langle \hat{S}_y \rangle_0^2)^{3/2}} e^{\frac{-7 \kappa r_n }{2 N}} + \mathcal{O}\left(\frac{\kappa^2}{\omega^2_0}\right),
%			\label{eq:34}
%		\end{equation}	   		 	
%	\end{widetext}
    The $\pm$ sign corresponds to the local maximum or minimum points of $\langle \hat{S}_{z} \rangle$, respectively. 
    The first term on the right-hand side represents the decay rate near the steady-state, while the second term is the exponentially decaying memory effect. 
    To better illustrate Eq.~(\ref{eq:34}), we consider a spin-squeezed coherent state generated via one-axis twisting,
    \begin{equation}
    	|\psi\rangle = e^{-i\chi t (\hat{S}_{y} - \hat{S}_{z})^2 } \, |\theta ,\varphi\rangle .
    \end{equation}
    Here, $e^{-i\chi t (\hat{S}_{y} - \hat{S}_{z})^2 }$ is the unitary evolution operator corresponding to one-axis twisting along the direction $\vec{n} = (0,1,-1)$, which induces spin squeezing and generates a nonzero correlation between $\hat{S}_y$ and $\hat{S}_z$. 
    And the spin coherent state $|\theta,\varphi\rangle$ is defined as~\cite{kitagawa1993squeezed}
    \begin{equation}
    	|\theta,\varphi\rangle = e^{-i\theta(\hat{S}_x \sin\varphi - \hat{S}_y \cos\varphi)} \, | S, S\rangle .
    \end{equation}
    In Fig.~\ref{fig3}, we take $\theta=\pi/3$, $\varphi=\pi/2$, and $\chi t = 0.2$. It demonstrates that our solution (Eq.~\ref{eq:34}) fits the simulated results of Eq.~\eqref{eq:1}.
	
	\section{SRWA with an Alternative Rotation Frequency And Error Analysis}
	\label{sec:new}	
	
	\begin{figure}[t]
		\renewcommand{\figurename}{FIG.}
		\centering
		
		% --- Row 1 ---
		\includegraphics[scale=0.18]{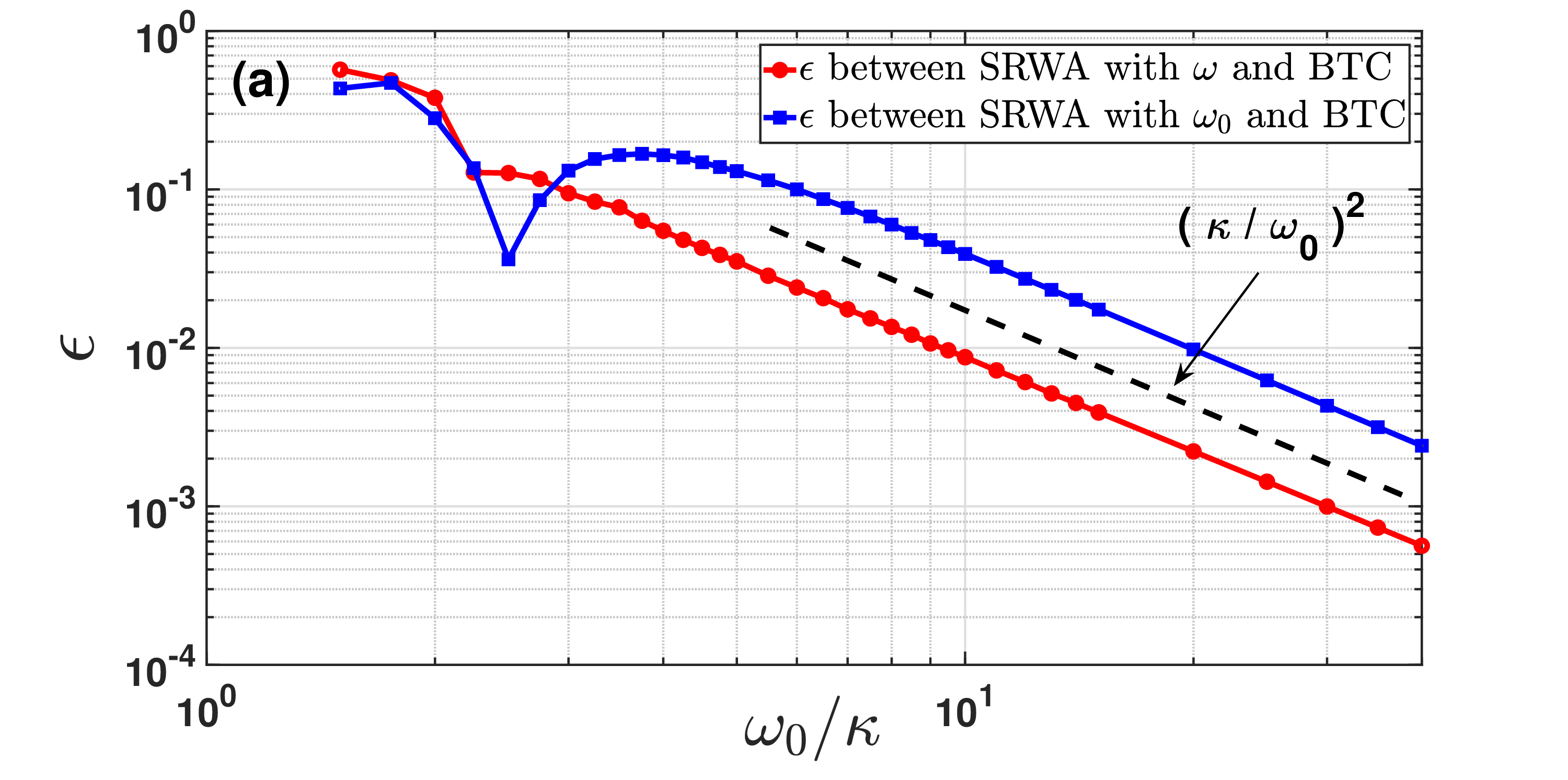}
		
		\vspace{6pt}
		
		% --- Row 2 ---
		\includegraphics[scale=0.18]{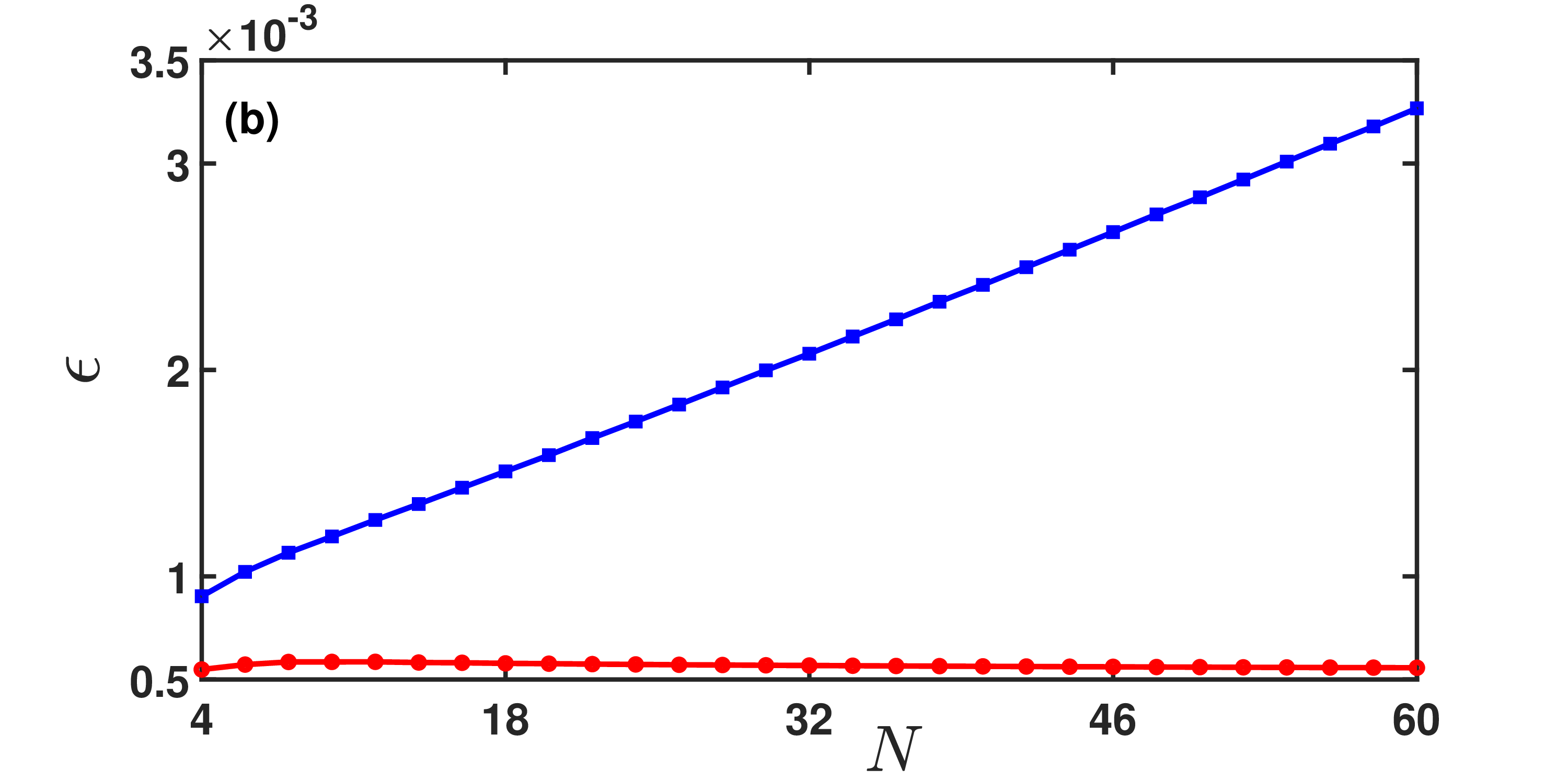}
		
		\vspace{6pt}

		\caption{Error analysis of the SRWA.
			The initial state is chosen as all spins pointing down.
			In panel (a), the red circles and blue squares show the absolute errors $\epsilon$ as functions of $\omega_0$, obtained from the SRWA with rotation frequencies $\omega$ and $\omega_0$, respectively.
     		In panel (b), the red circles and blue squares show the absolute errors $\epsilon$ as functions of system size $N$.}
		\label{fig6}
	\end{figure}

     In the above sections, the SRWA leads to a relative erorr of order
     $\left( \kappa / \omega_{0} \right)^{2}$.
     However, the dimensionless parameter $N$ also affects the accuracy of the results. For large $N$, the long-time predictions of the SRWA with the rotation frequency $\omega_{0}$ exhibit a pronounced increase in deviation. To improve the approximation in this regime, we discuss the SRWA with an alternative rotation frequency. According to Eq.~\eqref{eq:31}, the difference between the driving period $T_0$ and the oscillation period $T$
     leads to a difference between the driving frequency $\omega_0$ and the oscillation frequency $\omega$. Near the steady state, and upon neglecting the exponentially
     decaying terms, we obtain
	\begin{align}
		\omega 
		= \omega_{0} 
		- \frac{\kappa^{2}\!\left(4N^{2}+8N-11\right)}{8N^{2}\omega_{0}}
		+ \mathcal{O}\!\left(\frac{\kappa^2}{\omega_{0}^{2}}\right).
		\label{eq:35}
	\end{align}
	
	In this case, the corresponding effective Lindblad 
	superoperators $\bar{\mathcal{L}}^{(0)}_{c}$ and $\bar{\mathcal{L}}^{(1)}_{c}$, 
	as well as the effective quantum dynamical semigroup $\mathcal{G}_{c}(s)$, take the following forms
	\begin{align}
		\bar{\mathcal{L}}^{(0)}_{c} \hat{\rho}
		&=
		i (\omega - \omega_{0}) \left[\hat{S}_{x},\hat{\rho}\right]
		+ \bar{\mathcal{L}}^{(0)} \bigg|_{\omega_0 \to \omega} \hat{\rho},
		\\[6pt]
		\bar{\mathcal{L}}^{(1)}_{c} \hat{\rho}
		&=
		\bar{\mathcal{L}}^{(1)} \bigg|_{\omega_0 \to \omega} \hat{\rho}
		+ \mathcal{O}\!\left(\frac{\kappa^{2}}{\omega_{0}^{2}}\right),
		\\[6pt]
		\mathcal{G}_{c}(s) \hat{\rho}
		&=
		i (\omega - \omega_{0}) s \left[\hat{S}_{x},\hat{\rho}\right]
		+ \mathcal{G} (s) \bigg|_{\omega_0 \to \omega} \hat{\rho}.
	\end{align}

	To evaluate the accuracy of the SRWA, we consider the time-averaged absolute error of $\langle \hat{S}_z \rangle$ over one oscillation period around the relaxation time $\gamma^{-1} = 2N / (3\kappa)$ for the fully spin-down initial state (see Eq.~\ref{eq:n2}). 
	It is defined as
	\begin{equation}
		\epsilon
		= \frac{1}{T} \int_{\frac{1}{\gamma} - \frac{T}{2}}^{\frac{1}{\gamma} + \frac{T}{2}}
		\left| m_z(t) _{\mathrm{SRWA}} 
		- m_z(t) _{\mathrm{BTC}} \right| \, \mathrm{d}t,
		\label{eq:n3}
	\end{equation}
	where the subscripts indicate that the results are obtained by using the SRWA and the numerical simulation of  Eq.~\eqref{eq:1} (BTC).
	
	In Fig.~\ref{fig6}a, The red and blue curves correspond to the error $\epsilon$ of the SRWA with $\omega$ and $\omega_0$, respectively. 
	Both curves exhibit a power-law scaling $\epsilon \sim (\kappa / \omega_0)^2$, consistent with the above analysis.
	At low frequencies, however, both curves exhibit a turning point: the blue curve shows a bending, while the red curve develops a more pronounced kink. 
	This transition reflects the rapid growth of the SRWA error in this regime, which leads to a noticeable distortion of the oscillatory behavior.
	
	In Fig.~\ref{fig6}b, we further investigate the dependence of $\epsilon$ on system size $N$. 
	The blue curve, corresponding to the SRWA with $\omega_0$, increases with $N$. Meanwhile the red curve, corresponding to the SRWA with $\omega$, remains nearly constant. 
	This indicates that $\epsilon \propto N (\kappa / \omega_0)^2$ for the SRWA with $\omega_0$, while $\epsilon \propto (\kappa / \omega_0)^2$ for the SRWA with $\omega$. 
	Therefore for large $N$, choosing an appropriate rotation frequency effectively suppresses the accumulation of errors with increasing system size, leading to an improvement in the accuracy of the long-time dynamics.	
	
	\section{Discussion and Conclusion}
	\label{sec:discussion and conclusion}
	% 深入物理解读、与文献对比、方法适用性与局限	% 总结主要发现，展望未来工作
	In the BTC phase, the long relaxation time leads to the accumulation of error in the mean-field approximation, preventing it from providing a reliable description of system's long-time evolution. In contrast, the SRWA yields reliable full-time dynamics, as we employ two distinct effective descriptions to capture the short-time and long-time evolution, respectively, thereby effectively suppressing the accumulation of errors beyond the interval between adjacent stroboscopic points. This approach not only captures the decay of the order parameter, but also ensures the validity of the approximation in the long-time regime. Furthermore, we provide a quantitative error analysis for the SRWA as a function of driving frequency $\omega_{0}$ and system size $N$.
	
	Based on the SRWA, we find that the competition among dephasing processes along three directions leads to persistent oscillations, which signal the onset of the BTC phase. Beyond constructing the effective dynamics, our approach also yields analytical expressions for the steady-state density operator, the oscillation period, and the decay rate of observables in the regime where the coherent energy splitting exceeds the dissipation rate. In addition to the evolution of $\langle \hat{S}_z \rangle$, we further derive analytical expressions for other observables, including but not limited to $\langle \hat{S}_x \rangle$, $\langle \hat{S}_y \rangle$, $\langle \hat{S}_x^2 \rangle$, $\langle \hat{S}_y^2 \rangle$, and $\langle \hat{S}_z^2 \rangle$, as presented in Appendix~\ref{appendix:SRWA}.
	 
	Besides BTCs, our approach can also be extended to study the formation of time crystals in other periodically driven open quantum systems, such as Dicke models~\cite{gong2018discrete,zhu2019dicke}, spin-star models~\cite{krishna2023measurement} and in a single-mode nonlinear cavity~\cite{li2024time}.
	In particular, the evolution in cooperative fluorescence systems~\cite{drummond1978volterra,carmichael1980analytical,PhysRevLett.121.035301}, also called a driven-dissipative Dicke model in Ref.~\cite{ferioli2023non}, has essentially the same structure as the BTC model, with the difference that the dissipation rate does not scale as \(1/N\). As a result, while it does not realize a genuine BTC phase (the envelope still decays even in the limit \(N \to \infty\)), it belongs to the same universality class as BTCs. 
	Despite this, the applicability of the SRWA does not depend on the system size \(N\) and therefore remains applicable in the finite-size regime. Experimentally, the effective atom number is about 10, placing the system in such a regime. This enables a direct experimental verification of our theoretical predictions within the driven-dissipative Dicke model platform, such as the steady-state density matrix, the oscillation period, and the decay rate.
	
	\section*{Data availability}
	
	The numerical data supporting the findings of this work are available in the Zenodo repository, Ref.~\cite{Liu2026BTCdataset}.

	\begin{acknowledgments}
		This work was supported by Quantum Science and Technology-National Science and Technology Major Project (Grant No.~2024ZD0301000),
		Science Challenge Project (Grant No.~TZ2025017), 
		Science Foundation of Zhejiang Sci-Tech University (Grants No.~23062088-Y and No.~23062181-Y),
		and National Natural Science Foundation of China (Grant No.~12405046).
	\end{acknowledgments}
	%\vspace{-1em} % 适当调节
	\appendix
	\section{Master equation in the rotating frame}
	\label{appendix:rotating}
	
	The time-dependent ladder operators in the rotating frame are defined as  
	\begin{align}
		\tilde{S}_{\pm} &= \hat{U}^{\dagger} \hat{S}_{\pm} \hat{U}  \notag \\
		&= \hat{U}^{\dagger} \hat{S}_x \hat{U} \pm i \hat{U}^{\dagger} \hat{S}_y \hat{U} \notag \\
		&= \hat{S}_x \pm i \left[ \hat{S}_y \cos{(\omega_0 t)} - \hat{S}_z \sin{(\omega_0 t)} \right].
		\label{eq:a1}
	\end{align}			
	The time derivative of $\tilde{\rho}$ is given by  
	\begin{align}
		\frac{\mathrm{d}\tilde{\rho}}{\mathrm{d}t} &= \frac{\mathrm{d} \hat{U}^{\dagger}}{\mathrm{d}t} \hat{\rho} \hat{U} + \hat{U}^{\dagger}\hat{\rho} \frac{\mathrm{d}\hat{U}}{\mathrm{d}t} + \hat{U}^{\dagger} \frac{\mathrm{d}\hat{\rho}}{\mathrm{d}t} \hat{U} \notag \\
		&= i \hat{H} \tilde{\rho} - i \tilde{\rho} \hat{H} + \hat{U}^{\dagger} \frac{\mathrm{d}\hat{\rho}}{\mathrm{d}t} \hat{U} \notag \\
		&= i \omega_{0} \left[ \hat{S}_x, \tilde{\rho} \right] + \hat{U}^{\dagger} \frac{\mathrm{d}\hat{\rho}}{\mathrm{d}t} \hat{U}.
		\label{eq:a2}
	\end{align}	
	Therefore, the master equation in the rotating frame, which comes from Eq.~\eqref{eq:1}, takes the following form
	\begin{equation}
		\frac{\mathrm{d} \tilde{\rho}}{\mathrm{d} t} = \frac{ \kappa}{N} \left( 2 \tilde{S}_- \tilde{\rho} \tilde{S}_+ - \tilde{S}_+ \tilde{S}_- \tilde{\rho} -  \tilde{\rho} \tilde{S}_+ \tilde{S}_- \right).
		\label{eq:a3}
	\end{equation}
	Substituting Eq.~\eqref{eq:a1} into Eq.~\eqref{eq:a3}, we have
	\begin{widetext}
	\begin{equation}
		\begin{aligned}
			\frac{\mathrm{d} \tilde{\rho}}{\mathrm{d} t} = \frac{\kappa}{N} \bigg\{ 
			& \left(2\hat{S}_x \tilde{\rho} \hat{S}_x -  \hat{S}_x ^2 \tilde{\rho} -  \tilde{\rho} \hat{S}_x ^2 \right)  + \left( \hat{S}_y \tilde{\rho} \hat{S}_y - \frac{1}{2} \hat{S}_y ^2 \tilde{\rho} - \frac{1}{2} \tilde{\rho} \hat{S}_y ^2 \right)  + \left( \hat{S}_z \tilde{\rho} \hat{S}_z - \frac{1}{2} \hat{S}_z ^2 \tilde{\rho} - \frac{1}{2} \tilde{\rho} \hat{S}_z ^2 \right) \\
			& + i\cos(\omega_0 t) \bigg[ \left( 2\hat{S}_x \tilde{\rho} \hat{S}_y -  \hat{S}_y \hat{S}_x \tilde{\rho} -  \tilde{\rho} \hat{S}_y \hat{S}_x \right)  -\left( 2\hat{S}_y \tilde{\rho} \hat{S}_x - \hat{S}_x \hat{S}_y \tilde{\rho} - \tilde{\rho} \hat{S}_x \hat{S}_y \right) \bigg]\\
			& + i\sin(\omega_0 t) \bigg[ \left( 2\hat{S}_z \tilde{\rho} \hat{S}_x -  \hat{S}_x \hat{S}_z \tilde{\rho} -  \tilde{\rho} \hat{S}_x \hat{S}_z \right)  -\left( 2\hat{S}_x \tilde{\rho} \hat{S}_z - \hat{S}_z \hat{S}_x  \tilde{\rho} - \tilde{\rho} \hat{S}_z \hat{S}_x \right)  \bigg]\\
			& + \cos( 2\omega_0 t) \bigg[ \left( \hat{S}_y \tilde{\rho} \hat{S}_y - \frac{1}{2} \hat{S}_y ^2 \tilde{\rho} - \frac{1}{2} \tilde{\rho} \hat{S}_y ^2 \right)  -\left( \hat{S}_z \tilde{\rho} \hat{S}_z - \frac{1}{2} \hat{S}_z ^2 \tilde{\rho} - \frac{1}{2} \tilde{\rho} \hat{S}_z ^2 \right) \bigg]\\
			& - \sin( 2\omega_0 t) \bigg[ \left(  \hat{S}_y \tilde{\rho} \hat{S}_z - \frac{1}{2} \hat{S}_z \hat{S}_y \tilde{\rho} - \frac{1}{2} \tilde{\rho} \hat{S}_z \hat{S}_y \right)  + \left( \hat{S}_z \tilde{\rho} \hat{S}_y - \frac{1}{2} \hat{S}_y \hat{S}_z \tilde{\rho} - \frac{1}{2} \tilde{\rho} \hat{S}_y \hat{S}_z \right) \bigg] \bigg\}.
		\end{aligned}
		\label{eq:a4}
	\end{equation}
	\end{widetext}
	Under the RWA, the contributions of the time-dependent oscillating terms vanish in the limit $\omega_0 \to \infty$ and we obtain
	\begin{equation}
		\begin{aligned}
			\frac{\mathrm{d} \tilde{\rho}}{\mathrm{d} t} &= \frac{\kappa}{N} \bigg( 
			2\hat{S}_x \tilde{\rho} \hat{S}_x -  \hat{S}_x ^2 \tilde{\rho} -  \tilde{\rho} \hat{S}_x ^2 \\
			& \quad +  \hat{S}_y \tilde{\rho} \hat{S}_y - \frac{1}{2} \hat{S}_y ^2 \tilde{\rho} - \frac{1}{2} \tilde{\rho} \hat{S}_y ^2 \\
			& \quad + \hat{S}_z \tilde{\rho} \hat{S}_z - \frac{1}{2} \hat{S}_z ^2 \tilde{\rho} - \frac{1}{2} \tilde{\rho} \hat{S}_z ^2 
			\bigg).
		\end{aligned}
		\label{eq:a5}
	\end{equation}

	\section{Derivation of the Effective Lindblad Superoperators: 
		\(\bar{\mathcal{L}}^{(0)}\) and \(\bar{\mathcal{L}}^{(1)}\), and the Effective Quantum Dynamical Semigroup}
	\label{appendix:longtime and shorttime}
	
	We define the superoperators $\hat{O}_{L}$ ($\hat{O}_{R}$) to represent the operator acting on the density matrix from the left (right) as follows	
	\begin{equation}
		\hat{O}_{L}\,\hat{\rho} = \hat{O}\hat{\rho} 
		\qquad 
		\hat{O}_{R}\,\hat{\rho} = \hat{\rho} \hat{O}.
		\label{eq:b1}
	\end{equation}
	By applying the commutator on the density matrix, $[A, B]\rho \equiv AB\rho - BA\rho$, we obtain
	\begin{align}
		\left[ A^L, B^L\right] &= \left[ A, B \right]^L \notag \\
		\left[ A^R, B^R\right] &= - \left[ A, B \right]^R \notag \\
		\left[ A^L, B^R\right] &= \left[ A^R, B^L\right] = 0
		\label{eq:b2}
	\end{align}
	Then, the Lindblad superoperator in Eq.~(4) is expressed as
	\begin{align}
		\mathcal{L} = \frac{ \kappa}{N} \left( 
		2\tilde{S}_{-}^{L} \tilde{S}_{+}^{R} 
		-  \tilde{S}_{+}^{L} \tilde{S}_{-}^{L} 
		-  \tilde{S}_{-}^{R} \tilde{S}_{+}^{R} 
		\right).
		\label{eq:b3}
	\end{align}	
	By substituting Eqs.~\eqref{eq:a1} and~\eqref{eq:b2} into Eqs.~\eqref{eq:10a}, ~\eqref{eq:10b} and performing the integration, we obtain
	\begin{align}
		\bar{\mathcal{L}}^{(0)} 		
		&= \frac{ \kappa}{N} \bigg[
		2\hat{S}_{x}^{L} \hat{S}_{x}^{R} - \left(\hat{S}_{x}^{L}\right) ^2 -  \left(\hat{S}_{x}^{R}\right) ^2 \nonumber \\
		&\quad +  \hat{S}_{y}^{L} \hat{S}_{y}^{R} - \frac{1}{2} \left(\hat{S}_{y}^{L}\right) ^2 - \frac{1}{2} \left(\hat{S}_{y}^{R}\right) ^2 \nonumber \\
		&\quad +  \hat{S}_{z}^{L} \hat{S}_{z}^{R} - \frac{1}{2}	\left(\hat{S}_{z}^{L}\right) ^2 - \frac{1}{2} \left(\hat{S}_{z}^{R}\right) ^2
		\bigg],
		\label{eq:b4}
	\end{align}
		\begin{widetext}
		\begin{align}
			\bar{\mathcal{L}}^{(1)} \tilde{\rho} =
			& \frac{i \kappa^2}{8 \omega_{0} N^2} \bigg[ 
			16 \left( \hat{S}_x ^2 \tilde{\rho} \hat{S}_x - \hat{S}_x \tilde{\rho} \hat{S}_x ^2 \right) -5 \left( \tilde{\rho} \hat{S}_x - \hat{S}_x \tilde{\rho} \right)  + 10 \left( \hat{S}_x \hat{S}_y \tilde{\rho} \hat{S}_y - \hat{S}_y \tilde{\rho} \hat{S}_y \hat{S}_x \right)  + 18 \left( \hat{S}_x \hat{S}_z \tilde{\rho} \hat{S}_z - \hat{S}_z \tilde{\rho} \hat{S}_z \hat{S}_x \right)  \notag\\
			& + 6 \left( \hat{S}_x \tilde{\rho} \hat{S}_y ^2 + \hat{S}_y \tilde{\rho} \hat{S}_x \hat{S}_y - \hat{S}_y \hat{S}_x \tilde{\rho} \hat{S}_y - \hat{S}_y ^2 \tilde{\rho} \hat{S}_x \right)  + 2 \left(\hat{S}_z \hat{S}_x \tilde{\rho} \hat{S}_z + \hat{S}_z ^2 \tilde{\rho} \hat{S}_x - \hat{S}_z \tilde{\rho} \hat{S}_x \hat{S}_z - \hat{S}_x \tilde{\rho} \hat{S}_z ^2 \right) \notag \\ 
			& + 2 \left(\tilde{\rho} \hat{S}_z \hat{S}_x \hat{S}_z - \hat{S}_z \hat{S}_x \hat{S}_z \tilde{\rho}\right)  - 6 \left( \tilde{\rho} \hat{S}_y \hat{S}_x \hat{S}_y - \hat{S}_y \hat{S}_x \hat{S}_y \tilde{\rho} \right)  - 4 \left( \tilde{\rho} \hat{S}_x \hat{S}_z + \tilde{\rho} \hat{S}_z \hat{S}_x - \hat{S}_x \hat{S}_z \tilde{\rho} - \hat{S}_z \hat{S}_x \tilde{\rho} \right) 
			\bigg] \notag \\ 
			& + \frac{2 \kappa^2}{\omega_{0} N^2} \bigg[
			\left( \hat{S}_y \tilde{\rho} \hat{S}_z ^2 + \hat{S}_z ^2 \tilde{\rho} \hat{S}_y - \hat{S}_z \hat{S}_y \tilde{\rho} \hat{S}_z - \hat{S}_z \tilde{\rho} \hat{S}_y \hat{S}_z \right)  +2 \left( \hat{S}_y \tilde{\rho} \hat{S}_x ^2 + \hat{S}_x ^2 \tilde{\rho} \hat{S}_y - \hat{S}_x \hat{S}_y \tilde{\rho} \hat{S}_x - \hat{S}_x \tilde{\rho} \hat{S}_y \hat{S}_x \right) 
			\bigg].
			\label{eq:11}
		\end{align}
		\end{widetext}	
	\begin{widetext}
		Furthermore, by substituting Eqs.~\eqref{eq:a1} and~\eqref{eq:b2} into Eq.~\eqref{eq:12}, we obtain
	  \begin{align}
	  	\mathcal{G}(s)\tilde{\rho} 
	  	= &\ \frac{ \kappa s}{2N} 
	  	\Big[  \left( 4 \hat{S}_x \tilde{\rho} \hat{S}_x - 2\hat{S}_x^2 \tilde{\rho} - 2\tilde{\rho}\hat{S}_x^2 \right)  
	  	+  \left( 2\hat{S}_y \tilde{\rho}\hat{S}_y - \hat{S}_y^2 \tilde{\rho} - \tilde{\rho}\hat{S}_y^2 \right)  
	  	+  \left( 2\hat{S}_z \tilde{\rho}\hat{S}_z - \hat{S}_z^2 \tilde{\rho} - \tilde{\rho}\hat{S}_z^2 \right) \Big] \notag \\
	  	& + \frac{\kappa }{4\omega_{0} N} \Big[ 
	  	\sin\!\left( 2\omega_0 s \right) \left( 2\hat{S}_y \tilde{\rho}\hat{S}_y - \hat{S}_y^2 \tilde{\rho} - \tilde{\rho}\hat{S}_y^2 \right) 
	  	- \sin\!\left( 2\omega_0 s \right) \left( 2\hat{S}_z \tilde{\rho}\hat{S}_z - \hat{S}_z^2 \tilde{\rho} - \tilde{\rho}\hat{S}_z^2 \right) \notag \\
	  	& - 2 \sin^2\!\left( \omega_0 s \right) \left( 2\hat{S}_y \tilde{\rho}\hat{S}_z - \hat{S}_z \hat{S}_y \tilde{\rho} - \tilde{\rho}\hat{S}_z \hat{S}_y \right)  
	  	- 2 \sin^2\!\left( \omega_0 s \right) \left( 2\hat{S}_z \tilde{\rho}\hat{S}_y - \hat{S}_y \hat{S}_z \tilde{\rho} - \tilde{\rho}\hat{S}_y \hat{S}_z \right) \Big] \notag \\
	  	&+ \frac{i\kappa}{\omega_0 N} \Big[
	  	\sin\!\left( \omega_0 s \right) \left( 2\hat{S}_x \tilde{\rho}\hat{S}_y - \hat{S}_y \hat{S}_x \tilde{\rho} - \tilde{\rho}\hat{S}_y \hat{S}_x \right) 
	  	+ 2 \sin^2\!\left( \tfrac{\omega_0 s}{2} \right)
	  	\left( 2\hat{S}_z \tilde{\rho}\hat{S}_x - \hat{S}_x \hat{S}_z \tilde{\rho} - \tilde{\rho}\hat{S}_x \hat{S}_z \right) \notag \\
	  	& - \sin\!\left( \omega_0 s \right) \left( 2\hat{S}_y \tilde{\rho}\hat{S}_x - \hat{S}_x \hat{S}_y \tilde{\rho} - \tilde{\rho}\hat{S}_x \hat{S}_y \right) 
	  	- 2 \sin^2\!\left( \tfrac{\omega_0 s}{2} \right) 
	  	\left( 2\hat{S}_x \tilde{\rho}\hat{S}_z - \hat{S}_z \hat{S}_x \tilde{\rho} - \tilde{\rho}\hat{S}_z \hat{S}_x \right) \Big].
	  	\label{eq:13}
	  \end{align}	
	\end{widetext}

	\section{Expectation values of observables obtained by using the SRWA}
	\label{appendix:SRWA}
	
	Using $\bar{\mathcal{L}}^{(0)}$ and $\bar{\mathcal{L}}^{(1)}$ (Eqs.~\ref{eq:b2} and~\ref{eq:b3}), we obtain the effective  evolution equations for various observables along the stroboscopic points:
	
	\text{(1) Spin operators (to the order of $\kappa/\omega_{0}$)}
	\begin{align}
		\frac{\mathrm{d}\langle \hat{S}_{x} \rangle}{\mathrm{d} r_n} 
		&= -\frac{\kappa}{N}\,\langle \hat{S}_{x} \rangle 
		+ \frac{5\kappa^{2}}{2\omega_0 N^{2}} \,\langle \hat{S}_{x} \hat{S}_{y} + \hat{S}_{y} \hat{S}_{x} \rangle , 
		\label{eq:c1} \\[6pt]
		\frac{\mathrm{d}\langle \hat{S}_{y} \rangle}{\mathrm{d} r_n} 
		& = -\frac{3\kappa}{2N}\,\langle \hat{S}_{y} \rangle
		+ \frac{ \kappa^{2}}{2 \omega_0 N^{2}}
		\Bigg[ 
		\left( N^2 +2N - \frac{11}{4} \right) \langle \hat{S}_{z} \rangle \notag \\
		&\quad+ 2\langle \hat{S}_{x}^{2} - \hat{S}_{z}^{2} \rangle  + 12 \langle \hat{S}_{y}^{2} \rangle 
		\Bigg] ,
		\label{eq:c2} \\[6pt]
		\frac{\mathrm{d}  \langle \hat{S}_{z}  \rangle  }{\mathrm{d} r_n} 
		&= -\frac{3\kappa }{2 N} \langle \hat{S}_{z} \rangle - \frac{ \kappa^{2} }{2 \omega N^{2} } \Bigg[
		\left( N^2 +2N -\frac{11}{4}  \right) \langle \hat{S}_{y} \rangle  \notag \\
		&\quad - 7 \langle  \hat{S}_{y}\hat{S}_{z} + \hat{S}_{z}\hat{S}_{y} \rangle   
		\Bigg] , 
		\label{eq:c3} 
	\end{align}
	\text{(2) Square terms and cross term [to the order of $(\kappa/\omega_{0})^0$ ]}
	\begin{align}
		&\frac{\mathrm{d}\langle \hat{S}_{x}^{2} \rangle}{\mathrm{d} r_n} 
		 = \frac{\kappa}{N}\Big(-2\langle \hat{S}_{x}^{2} \rangle 
		+ \langle \hat{S}_{y}^{2} \rangle + \langle \hat{S}_{z}^{2} \rangle\Big) ,
		\label{eq:c4} \\[6pt]
		&\frac{\mathrm{d}\langle \hat{S}_{y}^{2} \rangle}{\mathrm{d} r_n} 
		= \frac{\kappa}{N}\Big(\langle \hat{S}_{x}^{2} \rangle 
		- 3\langle \hat{S}_{y}^{2} \rangle + 2\langle \hat{S}_{z}^{2} \rangle\Big) ,
		\label{eq:c5} \\[6pt]
		&\frac{\mathrm{d}\langle \hat{S}_{z}^{2} \rangle}{\mathrm{d} r_n} 
		= \frac{\kappa}{N}\Big(\langle \hat{S}_{x}^{2} \rangle 
		+ 2\langle \hat{S}_{y}^{2} \rangle - 3\langle \hat{S}_{z}^{2} \rangle\Big) ,
		\label{eq:c6} 	\\[6pt]
		&\frac{\mathrm{d}\langle \hat{S}_{y}\hat{S}_{z}+\hat{S}_{z}\hat{S}_{y} \rangle}{\mathrm{d} r_n} 
		= - \frac{5 \kappa}{N} \langle \hat{S}_{y}\hat{S}_{z}+\hat{S}_{z}\hat{S}_{y} \rangle .
		\label{eq:c7}
	\end{align}
	
	Solving the differential equations Eqs.~\eqref{eq:c1}–\eqref{eq:c7}, we obtain the following solutions:
	
	\text{(1) Spin operators}
	\begin{align}
	    \langle \hat{S}_{x}  \rangle (r_n) 
		& = \left \langle \hat{S}_{x} \right \rangle _{0}  e^{-\frac{ \kappa r_n}{N}} 
		\notag \\
		&\quad + \frac{5\kappa }{3\omega_0 N} \left \langle \hat{S}_{x} \hat{S}_{y} + \hat{S}_{y} \hat{S}_{x} \right \rangle _{0} \left( e^{-\frac{ \kappa r_n}{N}} - e^{-\frac{5 \kappa r_n}{2N}} \right)
		\label{eq:c8} \\
		\langle \hat{S}_{y} \rangle (r_n) 
		& =	\left \langle \hat{S}_{y} \right \rangle _{0}  e^{-\frac{3\kappa r_n }{2N}} \notag \\
		&\quad + \frac{ \kappa^{2}  }{2 \omega_0 N^{2} } 
		\left(N^2 +2N -\frac{11}{4}\right)  \left \langle \hat{S}_{z} \right \rangle _{0}  r_n  e^{-\frac{3 \kappa r_n}{2N}} \notag \\
		&\quad	+ \frac{\kappa (N+2) }{3\omega_0 }   \left( 1 - e^{-\frac{3 \kappa r_n}{2N}} \right) \notag \\
		&\quad - \frac{\kappa (N+2) }{12 \omega_0 }   \left( e^{-\frac{3 \kappa r_n}{N}} - e^{-\frac{3 \kappa r_n}{2N}} \right) \notag \\
		&\quad + \frac{\kappa  }{\omega_0 N} \left \langle  \hat{S}_{x}^{2} \right \rangle _{0}  \left( e^{-\frac{3 \kappa r_n}{N}} - e^{-\frac{3 \kappa r_n}{2N}} \right) \notag \\
		&\quad - \frac{\kappa }{\omega_0 N} \left \langle \hat{S}_{y}^{2} - \hat{S}_{z}^{2} \right \rangle _{0}  \left( e^{-\frac{5 \kappa r_n}{N}} - e^{-\frac{3 \kappa r_n}{2N}} \right) ,
		\label{eq:c9}  \\
		\langle \hat{S}_{z}  \rangle (r_n) 	        	
		&=   \langle \hat{S}_{z}  \rangle _{0}  e^{-\frac{3\kappa r_n }{2N}} \notag \\
		&\quad- \frac{ \kappa^{2}  }{2 \omega_0 N^{2} } 
		\left(
		N^2 +2N -\frac{11}{4} 
		\right) \left \langle \hat{S}_{y} \right \rangle _{0}       		        	 
		r_n  e^{-\frac{3 \kappa r_n}{2N}} \notag \\
		&\quad - \frac{\kappa }{\omega_0 N } \left \langle \hat{S}_{y} \hat{S}_{z} + \hat{S}_{z} \hat{S}_{y} \right \rangle _{0}  \left( e^{-\frac{5 \kappa r_n}{N}} - e^{-\frac{3 \kappa r_n}{2N}} \right) ,
		\label{eq:c10}
	\end{align}
	\text{(2) Square terms and cross term}
	\begin{align} 
		&\langle \hat{S}_{x}^{2} \rangle (r_n)
		= \frac{N(N+2)}{12}  
		- \left[ \frac{N(N+2)}{12} - \langle \hat{S}_{x}^{2} \rangle _{0}\right] 
		e^{-\frac{3\kappa r_n}{N}}, \label{eq:c11} \\
		&\langle \hat{S}_{y}^{2} \rangle (r_n) 
		= \frac{N(N+2)}{12} 
		+ \left[ \frac{N(N+2)}{24} - \frac{1}{2} \langle \hat{S}_{x}^{2} \rangle _{0} \right]
		e^{-\frac{3\kappa r_n}{N}} \notag \\
		&\qquad\qquad + \frac{1}{2}
		\langle \hat{S}_{y}^{2} -  \hat{S}_{z}^{2} \rangle _{0} 
		e^{-\frac{5\kappa r_n}{N}}, \label{eq:c12} \\
		&\langle \hat{S}_{z}^{2} \rangle (r_n) 
		= \frac{N(N+2)}{12}  
		+ \left[ \frac{N(N+2)}{24} - \frac{1}{2} \langle \hat{S}_{x}^{2} \rangle _{0}\right]  
		e^{-\frac{3\kappa r_n}{N}} \notag \\
		&\quad\qquad\quad - \frac{1}{2}
		\langle \hat{S}_{y}^{2} - \hat{S}_{z}^{2} \rangle _{0} 
	 	e^{-\frac{5\kappa r_n}{N}} ,   \label{eq:c13}  \\
		&\langle \hat{S}_y \hat{S}_z + \hat{S}_z \hat{S}_y \rangle (r_n) = \left \langle  \hat{S}_{y} \hat{S}_{z} + \hat{S}_{z} \hat{S}_{y}\right \rangle _{0}  e^{-\frac{5 \kappa r_n}{N}}, \label{eq:c14}
	\end{align}
	where $\langle \hat{O} \rangle_{0}$ represents the expectation value of the operator $\hat{O}$ in the initial state.

	The short-time evolution of the observables are given by the reduced quantum dynamical semigroup in Eq~\eqref{eq:13}:	   
	\begin{align}
		\langle \hat{S}_{x} \rangle (r_n, s)
		&= \left( 1 - \frac{ \kappa s}{N} \right) \langle \hat{S}_x \rangle (r_n)  \notag\\
		&\quad + \frac{ \kappa}{\omega_0 N} \Big[
		2\sin^2 \left( \frac{\omega_0 s}{2} \right)
		\langle \hat{S}_x \hat{S}_y + \hat{S}_y \hat{S}_x \rangle (r_n)   \notag \\
		&\quad +  \sin(\omega_0 s) 
		\langle \hat{S}_x \hat{S}_z + \hat{S}_z \hat{S}_x \rangle (r_n) \Big],
		\label{eq:c15}
	\end{align}
	\begin{align}
		&\langle \hat{S}_{y} \rangle (r_n, s) \notag \\
		& = \left(1- \frac{3 \kappa s }{2 N}\right)\Big[ \cos(\omega_0 s)\langle \hat{S}_y \rangle (r_n)  -\sin(\omega_0 s)\langle \hat{S}_z \rangle (r_n)  \Big] \notag\\
		& + \frac{\kappa}{2 \omega_0 N} \Big[ \sin(\omega_0 s) 
		\langle  \hat{S}_y \rangle (r_n)   \notag \\
		& +4  \langle \hat{S}_x^2 \rangle (r_n)  + 4\cos^2(\omega_0 s) \langle \hat{S}_z^2 \rangle (r_n)  \notag \\
		& +4 \sin^2(\omega_0 s) \langle \hat{S}_y^2 \rangle (r_n)   \notag \\
		& + 2 \sin(2\omega_0 s) 
		\langle \hat{S}_y \hat{S}_z + \hat{S}_z \hat{S}_y \rangle (r_n) \notag \\
		& - 4 \cos(\omega_0 s) 
		\langle \hat{S}_x^2 + \hat{S}_z^2 \rangle (r_n) \notag \\
		& - 2 \sin(\omega_0 s) 
		\langle \hat{S}_y \hat{S}_z + \hat{S}_z \hat{S}_y \rangle (r_n)    
		\Big].
		\label{eq:c16}
	\end{align}
	\begin{align}
		&\langle \hat{S}_{z} \rangle (r_n, s) \notag \\
		& = \left(1- \frac{3 \kappa s }{2 N}\right)
		\Big[ \cos(\omega_0 s)\langle \hat{S}_z \rangle (r_n) 
		+ \sin(\omega_0 s)\langle \hat{S}_y \rangle (r_n) \Big] \notag\\
		& + \frac{\kappa}{2 \omega_0 N} \Big[
		2 \sin(2\omega_0 s)
		\langle \hat{S}_z^2 - \hat{S}_y^2 \rangle (r_n) \notag \\
		& + 2 \cos(\omega_0 s) 
		\langle \hat{S}_y \hat{S}_z + \hat{S}_z \hat{S}_y \rangle (r_n) \notag \\
		&- \sin(\omega_0 s) \langle \hat{S}_z \rangle (r_n)  
		- 4 \sin(\omega_0 s) 
		\langle \hat{S}_x^2 + \hat{S}_z^2 \rangle (r_n) \notag \\
		& - 2 \cos(2\omega_0 s) 
		\langle \hat{S}_y \hat{S}_z + \hat{S}_z \hat{S}_y \rangle (r_n) \Big].
		\label{eq:c17}
	\end{align}
	
	The analytical solutions in Eqs.~\eqref{eq:c8}--\eqref{eq:c14} describe the decay of the observables at the stroboscopic points, 
	while Eqs.~\eqref{eq:c15}--\eqref{eq:c17} capture the oscillatory behavior between adjacent stroboscopic points. 
	Together, they provide an accurate description of the full-time dynamics of the BTC phase.

	\bibliographystyle{unsrt}   % 按引用顺序排序  % 选择参考文献格式，如 IEEE、plain、apalike 等
	\bibliography{ref}  % 这里写你的 .bib 文件名（不加 .bib 后缀）
	
\end{document}